\documentclass[12pt]{article}

\usepackage{epsf}

\newcommand{\be}{\begin{equation}}
\newcommand{\ee}{\end{equation}}
\newcommand{\bea}{\begin{eqnarray}}
\newcommand{\eea}{\end{eqnarray}}
\newcommand{\beq}{\begin{equation}}
\newcommand{\eeq}{\end{equation}}
\newcommand{\ba}{\begin{array}}
\newcommand{\ea}{\end{array}}
\newcommand{\beqa}{\begin{eqnarray}}
\newcommand{\eeqa}{\end{eqnarray}}

\newcommand{\cA}{{\cal A}}
\newcommand{\cO}{{\cal O}}

\newcommand{\no}{\nonumber}
\newcommand{\lsim}{\stackrel{<}{_\sim}}
\newcommand{\gsim}{\stackrel{>}{_\sim}}

\newcommand{\ccdot}{ \! \cdot \! }

\def\npb#1#2#3{    {\it Nucl. Phys. }{\bf B #1} (#2) #3}
\def\plb#1#2#3{    {\it Phys. Lett. }{\bf B #1} (#2) #3}

\def\prd#1#2#3{    {\it Phys. Rev. }{\bf D #1} (#2) #3}

\def\prl#1#2#3{    {\it Phys. Rev. Lett. }{\bf #1} (#2) #3}
\def\ptp#1#2#3{    {\it Prog. Theor. Phys. }{\bf #1} (#2) #3}

\def\ppnp#1#2#3{   {\it Prog. Part. Nucl. Phys. }{\bf #1} (#2) #3}
\def\rmp#1#2#3{    {\it Rev. Mod. Phys. }{\bf #1} (#2) #3}
\def\zpc#1#2#3{    {\it Z. Phys. }{\bf C #1} (#2) #3}
\def\ijmpa#1#2#3{  {\it Int. J. Mod. Phys. }{\bf A #1} (#2) #3}

\def\ibid#1#2#3{   {\it ibid. }{\bf #1} (#2) #3}
\def\jhep#1#2#3{   {\it JHEP  }{\bf #1} (#2) #3} 

\def\ncim#1#2#3{   {\it Nuovo Cim. }{\bf #1} (#2) #3}

\begin{document}

\thispagestyle{empty}
\begin{flushright}
LMU 12/03\\
INFNNA-IV-03-19 \\
July 2003
\end{flushright}
\vskip 3.0 true cm

\begin{center}
{\Large\bf Extracting short-distance physics \\ [5 pt]
from $K_{L,S} \rightarrow \pi^0 e^+ e^-$ decays} 
 \\ [25 pt]
{\sc Gerhard Buchalla${}^a$}, {\sc Giancarlo D'Ambrosio${}^b$},
and  {\sc Gino Isidori${}^{c}$} 
 \\ [25 pt]
{\sl ${}^a$Ludwig-Maximilians-Universit\"at M\"unchen, Sektion Physik, \\
     Theresienstr. 37, D-80333 M\"unchen, Germany} \\ [5 pt]
{\sl ${}^b$INFN, Sezione di Napoli and Dipartimento di Scienze Fisiche, \\ 
           Universit\`a di Napoli, I-80126 Napoli, Italy} \\ [5 pt]
{\sl ${}^c$INFN, Laboratori Nazionali di Frascati, I-00044 Frascati, Italy} 
 \\ [25 pt]
{\bf Abstract}

\end{center}


We present a new analysis of the rare decay $K_L\to\pi^0 e^+ e^-$
taking into account important experimental progress that has
recently been achieved in measuring $K_L\to\pi^0\gamma\gamma$ and 
$K_S\to\pi^0 e^+ e^-$. This includes a brief review of the
direct CP-violating component, a calculation of the indirect
CP-violating contribution, which is now possible after the measurement
of $K_S\to\pi^0 e^+ e^-$, and a re-analysis of the CP conserving part.
The latter is shown to be negligible, based on experimental input
from $K_L\to\pi^0\gamma\gamma$, a more general treatment of the
form factor entering the dispersive contribution, and on a comparison
with the CP violating rate, which can now be estimated reliably.
We predict $B(K_L\to\pi^0 e^+ e^-) = 
(3.2^{+1.2}_{-0.8})\times 10^{-11}$ in the
Standard Model, dominated by CP violation with a sizable contribution
$(\sim 40\%)$ from the direct effect, largely through interference
with the indirect one. Methods to deal with the severe backgrounds
for $K_L\to\pi^0 e^+ e^-$ using Dalitz-plot analysis and time-dependent
$K_L$--$K_S$ interference are also briefly discussed.

\def\thefootnote{\arabic{footnote}}
\setcounter{footnote}{0}
\setcounter{page}{0}


\newpage

\section{Introduction}
The flavour-changing 
neutral-current transition $K_{L} \rightarrow \pi ^{0}e^{+}e^{-}$ 
has been recognized since a long time 
to be one of the most interesting 
rare kaon decays. It shows 
an intriguing interplay of short and 
long distances, leading to a sum of comparable 
CP-conserving, direct- and indirect-CP-violating
contributions \cite{GW}. Until very recently, it was impossible
to estimate all these contributions with 
good accuracy, or to predict the total 
$K_{L} \rightarrow \pi ^{0}e^{+}e^{-}$ rate.
As a consequence, it was not clear to which extent 
this mode could be used as a probe of the 
Cabibbo-Kobayashi-Maskawa mechanism 
of CP violation. 
The situation has now been changed substantially by 
two new experimental results of the NA48 Collaboration:
the observation of the $K_S \rightarrow \pi ^{0}e^{+}e^{-}$ 
decay \cite{NA48KS} and the precise measurement of the $K_L \rightarrow \pi ^{0}\gamma\gamma$
spectrum at small diphoton-invariant mass \cite{NA48_KLpgg}.
The former allows us to evaluate the indirect-CP-violating part
of the amplitude, whereas the latter help us to estimate
the CP-conserving contribution. 
The purpose of this paper is a complete reanalysis of 
the $K_{L} \rightarrow \pi ^{0}e^{+}e^{-}$ decay
in view of this new experimental information.

The use of experimental data to estimate 
indirect-CP-violating and  CP-conserving contributions
of $K_{L} \rightarrow \pi ^{0}e^{+}e^{-}$ is not 
completely straightforward. The most delicate issue 
concerns the CP-conserving contribution and, 
particularly, the dispersive part of the   
$K_L \rightarrow \pi ^{0}\gamma^*\gamma^* \to \pi ^{0} e^{+}e^{-} $ 
amplitude. So far, this part of the amplitude has been 
estimated employing a specific model-dependent 
ansatz for the behavior of the $K_L \rightarrow \pi ^{0}\gamma^*\gamma^*$ 
vertex with off-shell photons \cite{DG}.
Here we generalize previous analyses using a more general 
parameterization of the $K_L \rightarrow \pi ^{0}\gamma^*\gamma^*$  
form factor, which satisfies both low- and high-energy constraints
and helps us to estimate the theoretical uncertainties in this calculation. 
Moreover, we show how to extract the information on
the on-shell $K_L \rightarrow \pi ^{0}\gamma\gamma$ amplitude 
relevant to  $K_{L} \rightarrow \pi ^{0}e^{+}e^{-}$
in a model-independent way, without relying on 
model-dependent assumptions on the former, such as 
the vector-meson-dominance parameterization in terms of 
$a_V$. As a result, we are able to 
derive a {\em conservative} upper bound on the 
total CP-conserving contribution of the 
$K_{L} \rightarrow \pi ^{0}e^{+}e^{-}$ rate,
which turns out to be well below the level 
of the interesting short-distance component.

As far as the CP-violating amplitude is concerned, 
the NA48 result on $B(K_S \rightarrow \pi ^{0}e^{+}e^{-})$ 
provides us with an unambiguous indication that the 
indirect-CP-violating contribution is large and
cannot be neglected. Here the most delicate issue 
is the model-dependent sign of the interference between 
direct-- and indirect-CP-violating components of the amplitude.
As we shall show, the measured value of  
$B(K_S \rightarrow \pi ^{0}e^{+}e^{-})$, 
together with theoretical arguments both of 
perturbative and non-perturbative nature, provides
a good indication in favour of a positive interference.
Following this indication, we predict 
$B(K_L \rightarrow \pi ^{0}e^{+}e^{-})_{\rm SM} \approx 3\times 10^{-11}$,
with a negligible CP-conserving component, about $40\%$ 
due to the clean short-distance direct-CP-violating amplitude
(mainly through the interference with the indirect-CP-violating 
one) and an overall theoretical error that to a large extent scales with 
the experimental error on $B(K_S \rightarrow \pi ^{0}e^{+}e^{-})$.

From a purely theoretical perspective we thus conclude that 
$K_L \rightarrow \pi ^{0}e^{+}e^{-}$ appears as one of the 
most interesting candidates for precision tests of CP violation 
in ${\Delta S=1}$ transitions. The most serious problem to 
reach this goal is the so-called Greenlee background 
\cite{Greenlee,KTeV_p0ee}, 
that is, the large irreducible experimental background 
induced by the decay $K_L \to \gamma \gamma e^+ e^-$.
The enhancement of the total $K_L \rightarrow \pi ^{0}e^{+}e^{-}$
rate due to the positive interference between 
direct-- and indirect-CP-violating amplitudes 
provides good news in this respect, suggesting that 
future high-statistics experiments could be able to 
detect the $K_L \rightarrow \pi ^{0}e^{+}e^{-}$ signal over 
Greenlee's background. As we shall show, a Dalitz-Plot analysis 
and, especially, time-dependent measurements could provide 
additional handles against this problem.

\medskip 

The paper is organized as follows: in Section~\ref{sect:short}
we briefly review the prediction for the 
short-distance direct-CP-violating amplitude of 
 $K_L \rightarrow \pi^0 e^+ e^-$ within the Standard Model (SM).
Section~\ref{sect:CPC} contains one of the two main results of this work, 
namely the new conservative estimate of the CP-conserving branching ratio.
The second main result is presented in Section~\ref{sect:CPmix}, where 
we analyse the interference between direct- and 
indirect-CP-violating amplitudes, and the prediction for
the total rate. A discussion of 
possible Dalitz-Plot and time-dependent analyses against 
the CP-conserving and, especially, Greenlee's background
is presented in Sections~\ref{sect:rates} and~\ref{sect:time}.
The results are summarized in the Conclusions.
The Appendix contains a self-contained model-independent 
analysis of the $K_L \rightarrow \pi ^{0}\gamma\gamma$ amplitude 
in chiral perturbation theory (CHPT) beyond the
lowest non-trivial order.

\section{Short-distance contribution to $K_L \rightarrow \pi^0 e^+ e^-$}
\label{sect:short}

The direct-CP-violating transition $K_2 \to \pi^0 (e^+ e^-)_{J=1}$,
where the lepton pair forms a vector or axial-vector state,
is dominated by short-distance dynamics and is calculable with 
high accuracy in 
perturbation theory \cite{GW,BLMM}. Following the notation of \cite{BBL},
the effective Hamiltonian necessary to compute this amplitude 
at next-to-leading order accuracy 
can be written as 
\beq
{\cal H}_{eff}^{|\Delta S| = 1} = \frac{G_F}{\sqrt{2}} \, 
V_{us}^* V_{ud} \Big[ \, \sum_{i=1}^{6,7V} ( z_i(\mu) + \tau y_i(\mu) )
Q_i(\mu) ~+~ \tau y_{7A}(M_W) Q_{7A}(M_W) \, \Big]\, +\, \mbox{h.c.}
\label{eq:heff}
\eeq
where $\tau = - (V_{ts}^* V_{td})/(V_{us}^* V_{ud})$ and 
$V_{ij}$ denote CKM matrix elements. Here $Q_{1,2}$
are the current--current operators,  
$Q_{3 \dots 6}$ the QCD penguin operators, and 
\begin{equation}
Q_{7V} \, =  \, \overline{s} \gamma^{\mu}(1-\gamma_5) d \, \overline{\ell} 
\gamma_{\mu} \ell~, \qquad \quad
Q_{7A} \, =  \, \overline{s} \gamma^{\mu} (1-\gamma_5) d \, \overline{\ell}
\gamma_{\mu} \gamma_5 \ell 
\label{eq:q7v7a}
\end{equation}
the leading electroweak operators.
Employing the standard CKM phase convention, the overall 
factor $V_{us}^* V_{ud}$
is real and direct CP violation is induced only by terms proportional to 
$\Im \tau $. Since $y_1=y_2\equiv 0$, only  $Q_{7V}$, $Q_{7A}$ and, 
in principle, the QCD penguin operators  $Q_{3 \dots 6}$,
are relevant to estimate the
$K_2 \to \pi^0 (e^+ e^-)_{J=1}$ amplitude.

Ignoring for the moment the contribution of $Q_{3 \dots 6}$, 
whose matrix elements vanish at the tree-level, one has
\beqa
&& A\left( K_L \rightarrow \pi e^+ e^- \right)_{\rm s.d.} ~=~ 
- \langle \pi(p_\pi)  e^+(k_1) e^-(k_2) \vert 
{\cal H}_{eff}^{|\Delta S| = 1} \vert  K_2(p) \rangle  \no\\
&& \qquad =~ i \frac{G_F}{\sqrt{2}} \Im (V_{ts}^* V_{td}) f_+(z) (p_\pi +p)^\mu
\left[ y_{7V}  \bar{u}(k_2) \gamma_\mu v(k_1) + y_{7A}  
\bar{u}(k_2) \gamma_\mu\gamma_5 v(k_1) \right]~. \quad 
\label{eq:SD_ampl}
\eeqa
where 
$f_+(z=q^2/m_K^2) \approx  1 + z (m_K/m_\rho)^2$ is the form factor of the 
vector current and $q^2= (p-p_\pi)^2$.
If considered alone, this theoretically clean part of the amplitude leads to
\beqa
B(K_L \to \pi^0 e^+ e^-)_{\rm CPV-dir}&=& \frac{\tau(K_L)}{\tau(K^+)} 
\frac{B(K^+_{e3})}{|V_{us}|^2}
( y_{7A}^2 + y_{7V}^2 ) \left[ \Im (V_{ts}^* V_{td}) \right]^2~, \no\\
&=& (2.4 \pm 0.2) \times 10^{-12}~\left[ \frac{\Im \lambda_t }{  10^{-4} } 
\right]^2~,
\label{eq:SD_BR}
\eeqa
where
\beq
\Im \lambda_t = \Im (V_{ts}^* V_{td}) \stackrel{ {}_{\rm SM}}{ \longrightarrow }
(1.36 \pm 0.12) \times 10^{-4}~ \cite{CKM} 
\eeq
and the numerical coefficient in (\ref{eq:SD_BR})  has been obtained 
using $V_{us}=0.2240 \pm 0.0036$ \cite{CKM}, $\alpha(M_Z)=1/129$,
$y_{7A} = -(0.68 \pm 0.03) \alpha(M_Z)$ and 
$y_{7V} = (0.73 \pm 0.04) \alpha(M_Z)$,
corresponding to ${\overline m}_t (m_t) = 167 \pm 5$~GeV and a renormalization
scale for $y_{7V}$ chosen between $0.8$ and $1.2$~GeV~\cite{BBL}.

The contribution of the QCD penguin operators cannot 
be estimated with good accuracy, due to unknown hadronic 
matrix elements. However, $Q_{3 \dots 6}$ are expected to yield  
a negligible correction to the leading electroweak operators:
\beq
\sum_{i=3}^6 y_i(\mu) \langle \pi  e^+  e^- \vert 
Q_i  \vert K \rangle \ll y_{7V}  \langle \pi  e^+  e^- \vert 
Q_{7V}  \vert K \rangle~.
\label{eq:assum_Q6}
\eeq
This assumption is strongly supported by
i) the corresponding relation for the quark-level matrix elements
\beq
\sum_{i=3}^6 y_i(\mu) \langle d  e^+  e^- \vert 
Q_i  \vert s \rangle \ll y_{7V}  \langle d e^+ e^- \vert 
Q_{7V}  \vert s \rangle~,
\label{eq:assum_Q6b}
\eeq
which can be checked explicitly in perturbation theory;
ii) the following order-of-magni\-tu\-de estimate for the impact 
of the $Q_i$ at the hadronic level:
\beq
\delta B(K_L \to \pi^0 e^+ e^-)_{Q_i} \sim \left( \frac{y_i}{ z_2 } \Im \tau \right)^2 
\frac{\tau(K_L)}{\tau(K^+)}  B(K^+ \to \pi^+ e^+ e^-) < 10^{-14}~.
\label{eq:assum_Q6c}
\eeq
Given these considerations, Eqs. (\ref{eq:SD_ampl}) and (\ref{eq:SD_BR})
can be considered as precise estimates of the 
$K_L \rightarrow \pi^0 e^+ e^-$
direct-CP-violating amplitude, and its corresponding branching ratio,
within the SM.

It is worth to recall that, being dominated by short distances, 
this part of the $K_L \rightarrow \pi^0 e^+ e^-$ amplitude is strongly 
sensitive to possible non-standard contributions. In particular, 
axial and vector components of the amplitude could separately 
or both be enhanced by factors of $\cO(1)$ with respect 
to the SM case. In optimistic but still realistic supersymmetric 
scenarios, the short-distance branching ratio in Eq.~(\ref{eq:SD_BR}) could 
exceed the $10^{-11}$ level \cite{BCIRS}.

\section{Estimates of the CP-conserving amplitude}
\label{sect:CPC}

The $| \pi^0 e^+ e^- \rangle$ state is 
not, in general, a CP eigenstate: 
its transformation under CP depends on the
angular momentum of the lepton pair. 
The short-distance Hamiltonian in Eq.~(\ref{eq:heff}) 
creates the lepton pair in a state 
of total angular momentum $J=1$ (and orbital angular momentum $L=0$) 
so that $| \pi^0 e^+ e^- \rangle_{\rm s.d.}$
has opposite CP with respect to $|K_2\rangle$.
On the contrary, the long-distance process 
$K_2 \rightarrow \pi ^{0}\gamma\gamma \to \pi^0 e^+ e^-$
can lead to final states with even $J$, allowed in the
limit of exact CP symmetry \cite{DG,Se88,EPR88,FR89,Se93,CDM93,CEP93}. 
The two-photon  exchange is not the only source 
of CP-conserving contributions to $K_L \rightarrow \pi^0 e^+ e^-$, 
in principle also dimension-8 operators generated by $W$-box diagrams 
can lead to a final state with $J\not=1$; however, 
such contributions turn out to be completely negligible \cite{BI98}.

In the limit of exact CP symmetry, the most general
amplitude describing the $K_{L}(p)\rightarrow \pi ^{0}(p_{3})\gamma 
(q_{1},\epsilon_{1})\gamma (q_{2},\epsilon _{2})$ transition,
with on-shell or off-shell photons, can be written as
\beqa
  && \cA( K_{L}\rightarrow \pi ^{0}\gamma \gamma ) =  
   \frac{G_{8} \alpha }{ 4\pi }
   \epsilon_{1 \mu} \epsilon_{2 \nu} 
   \Big[  A(z,y; q_1^2, q_2^2) 
    (q_{2}^{\mu}q_{1}^{\nu }-q_{1}\ccdot q_{2}~ g^{\mu \nu } )  \no \\
 && \qquad + 
   \frac{2B(z,y; q_1^2, q_2^2)}{m_{K}^{2}} 
    (p \ccdot q_{1}~ q_{2}^{\mu }p^{\nu} + p\ccdot q_{2} ~p^{\mu }q_{1}^{\nu}
   -p\ccdot q_{1}~ p\ccdot q_{2}~ g^{\mu \nu } 
   - q_{1}\ccdot q_{2}~ p^{\mu }p^{\nu } ) \Big]~, \quad
\label{eq:mmunu}
\eeqa
where 
\beq
y = \frac{ p\ccdot (q_{1}-q_{2})}{m_{K}^{2}}~, \qquad 
z = \frac{ (q_{1}+q_{2})^{2} }{ m_{K}^{2}}~, \qquad 
|G_{8}| = 9.0 \times 10^{-6}~{\rm GeV}^{-2}~.
\label{eq:yz}
\eeq
and $G_8=G_F |V_{us} V_{ud}| g_8/\sqrt{2}$ with $|g_8|=5.1$.
Due to Bose symmetry the invariant amplitudes 
$A$ and $B$ are symmetric under the exchange 
$q_{1} \leftrightarrow q_{2}$. 

Restricting the attention to the on-shell case ($q_1^2=q_2^2=0$),
the allowed range of the dimensionless variables is
\beq
 0\leq |y| \leq \frac{1}{2} \lambda^{1/2}(1,r_{\pi }^{2},z)~, \qquad 
 0\leq z\leq (1-r_{\pi})^{2}~,
\eeq
where 
\beq
\lambda (a,b,c) = a^{2}+b^{2}+c^{2}-2(ab+ac+bc)~, \qquad 
r_{\pi } = \frac{\displaystyle m_{\pi }}{\displaystyle m_{K}}~,
\eeq
and the unpolarized differential rate reads
\beq
\frac{\partial^{2}\Gamma }{\partial y \partial z}  = 
\frac{ G_8^2 \alpha^2 m^5_K } { 2^{13} \pi ^{5} }\left[  z^{2} 
| A + B |^{2}  + \left( y^{2} - 
  \frac{1}{4}\lambda (1,r_{\pi }^{2},z)\right)^{2}
| B |^{2}\right]~. 
\label{eq:doudif}
\eeq
In order to obtain the total rate from (\ref{eq:doudif}), the integration is
to be performed over positive values of $y$ only.

Instead of using the $A$--$B$ basis, one can employ 
the $S$--$B$ amplitude basis, where $S\equiv A+B$.
As can be seen from Eq.~(\ref{eq:doudif}), in the $S$--$B$ basis
the two amplitudes do not interfere in the on-shell unpolarized rate.
This indicates that $S$ and $B$  
describe transitions to different final states. 
Indeed, neglecting the $y$ dependence, 
$S$ describes the decay into $|\gamma\gamma\rangle_{J=0}$  
and $B$ the decay into $|\gamma\gamma\rangle_{J=2}$.
Since the transition  $|\gamma\gamma\rangle_{J=0} 
\to |e^+ e^-\rangle$ is forbidden in the limit $m_e \to 0$, 
if $S$ does not depend on $y$, only the $B$ amplitude can induce 
a non-negligible CP-conserving contribution to $K_L \to \pi^0 e^+ e^-$ 
\cite{Se88,EPR88,FR89}. 
The possible $y$ dependence of $S$ is strongly suppressed by 
the combined action of Bose symmetry, which forbids linear terms,
and chiral symmetry. For this reason we shall concentrate in the following 
mainly on the CP-conserving contribution to $K_L \to \pi^0 e^+ e^-$ 
generated by the $B$ amplitude.

Within CHPT the first non-vanishing contribution to 
the on-shell $K_{L}\rightarrow \pi ^{0}\gamma \gamma$ 
transition is generated at  $\cO(p^{4})$. 
At this order only a $y$-independent $S$ amplitude 
turns out to be different from zero \cite{EPR87}. However,
this parameter-free prediction underestimates the 
observed $K_{L}\rightarrow \pi^{0}\gamma \gamma$ rate 
roughly by a factor of 2, indicating the presence of 
sizable $\cO(p^{6})$ contributions. 
These have been widely discussed in the literature 
\cite{Se93,CDM93,CEP93,EPR90,KH94,DP97} and include both 
local counterterms (presumably generated by resonance 
exchange) and non-local terms due to unitarity corrections.
By power counting, $\cO(p^{6})$ local counterterms 
do not induce $y$-dependent terms in $S$ and $B$.
Unitarity corrections do induce a dependence on $y^2$ in $S$;
however, this is numerically rather suppressed \cite{CDM93,CEP93}.

\begin{figure}[t]
\begin{center}
\leavevmode
\epsfxsize=10.0 cm
\epsfbox{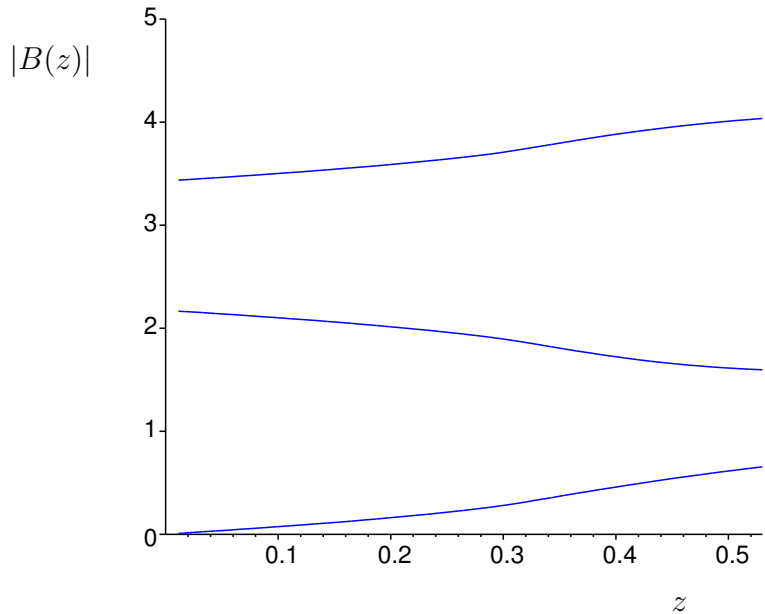}
\end{center}
\vskip -8.0 cm
\hskip  2.2 cm $|B(z)|$
\vskip  6.7 cm
\hskip  11.0 cm $z$
\vskip  0.3 cm
\caption{Absolute value of the $B$ amplitude for different values 
of the $\cO(p^6)$ local counterterms (see the Appendix).}
\label{fig:B}
\end{figure}

An updated discussion about the structure of the 
$K_{L}\rightarrow \pi^{0}\gamma \gamma$ amplitude 
beyond $\cO(p^{4})$ can be found in the Appendix. 
As far as the $B$ amplitude is concerned, 
here we simply note that $B(z)$ is well 
approximated by $B(0)$, whose value depended 
on a specific combination of $\cO(p^{6})$ counterterms.
The most efficient and model-independent strategy to 
determine the magnitude of $|B(0)|$ is provided by a measurement of 
$B(K_{L}\rightarrow \pi ^{0} \gamma \gamma )$ in the low diphoton invariant 
mass region. Neglecting the kinematical dependence of $S$ and $B$ 
(expected to be very mild in both cases below the $\pi\pi$ threshold) 
and setting the cut  $M_{\gamma \gamma} < 110$~MeV, to make contact with the 
NA48 analysis \cite{NA48_KLpgg}, we find
\beqa  
&& B(K_L \to \pi^0 \gamma\gamma)_{M_{\gamma \gamma} < 110~{\rm MeV}} \no\\
&& \qquad  = 2.1 \times 10^{-6} \times
 ~ \int_{z < 0.049} dz  
\left[  z^{2}    \lambda(1,r_{\pi }^{2},z)^{1/2}
| S(0) |^{2}  + \frac{1}{30} \lambda(1,r_{\pi }^{2},z)^{5/2}
| B(0) |^{2}\right] \quad  \no\\
&& \qquad = 2.0 \times 10^{-9} \times | B(0) |^{2} \times 
\left[ 1 + 0.04  \frac{|S(0)|^2}{|B(0)|^2}  \right]~.
\label{eq:B110}
\eeqa
If $S(0)/B(0)= \cO(1)$,
the experimental ratio
\beq
R_B = \frac{ B(K_L \to \pi^0 \gamma\gamma)_{M_{\gamma \gamma} < 110~{\rm MeV}}}
                     {2.0 \times 10^{-9}}
\label{eq:B111}
\eeq
provides an excellent approximation to $|B(0)|^2$; moreover, 
the inequality $|B(0)|^2 \leq R_B$ holds independently from any 
assumption on $S(0)/B(0)$. As shown in Fig.~\ref{fig:B},
the approximation $B(z)\approx B(0)$ works very well for 
$|B(0)| \gsim 1$, which is the case of interest for 
$K_{L}\rightarrow \pi ^{0} e^+e^-$.
Thus $R_B$ provides a
powerful model-independent tool to estimate (or constrain) the size of  
$|B(z)|$ in the whole physical range.

\begin{figure}[t]
\begin{center}
\leavevmode
\epsfysize=5 cm
\epsfbox{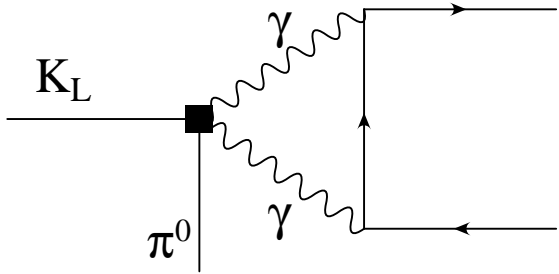}
\end{center}
\caption{Two-photon contribution to the $K_L \to \pi^0 e^+ e^-$
amplitude.}
\label{fig:loop}
\end{figure}

Having determined the absolute value of the $B$ amplitude
at $q_1^2=q_2^2=0$ from 
ex\-pe\-ri\-ments, we can use it as a (real) effective coupling 
for the $K_L \to \pi^0\gamma(q_1)\gamma(q_2)$ vertex 
in Fig.~\ref{fig:loop}, 
in order to estimate the CP-conserving  $K_L \to \pi^0 e^+ e^-$
amplitude. If we ignore the dependence on $q_{1,2}^2$
the two-photon dispersive 
integral turns out to be logarithmically divergent. 
To regularize it we employ the following ansatz 
\beq
B(z,y; q_1^2, q_2^2) = B(z) \times f(q_{1}^{2},q_{2}^{2})\
\label{eq:Boff}
\eeq
where the form factor 
\beq
f(q_{1}^{2},q_{2}^{2})=1+ a \left( \frac{\displaystyle q_{1}^{2}}{
\displaystyle q_{1}^{2}-m_{V}^{2}}+\frac{\displaystyle q_{2}^{2}}{%
\displaystyle q_{2}^{2}-m_{V}^{2}}\right) +b \frac{\displaystyle %
q_{1}^{2}q_{2}^{2}}{\displaystyle (q_{1}^{2}-m_{V}^{2})(q_{2}^{2}-m_{V}^{2})}~,
\label{eq:fq1q2}
\eeq
is defined in analogy with the analysis 
of the $K_{L} \to \gamma \gamma \to \mu^+\mu^-$
amplitude presented in \cite{DIP98}. 
This structure is dictated by
the assumption that VMD plays a crucial role in the matching between 
short and long distances (in the numerical analysis $m_V$ is
identified with $m_\rho \simeq 770$~MeV).
In order to obtain an ultraviolet convergent 
integral, as it is the case within the full theory,  we need to impose 
the condition 
\beq
1+2 a + b = 0~.  \label{eq:sumrule}
\eeq
In this way we are left with a single free parameter, that we can 
vary to estimate the theoretical uncertainties of this 
approach.\footnote{~In principle it could be possible to obtain 
additional constraints on this form-factor  
by means of $K_{L}\rightarrow \pi^{0}l^{+}l^{-}\gamma$ data \cite{DG_pgee}; 
however, the present experimental information is 
not accurate enough to extract any significant  constraint
\cite{KTeV_pgee}. }
Obviously Eq.~(\ref{eq:Boff}) does not represent 
the most general parameterization of the $B$ amplitude off shell.
However, we are not interested in the detailed 
structure of this amplitude, rather in its 
weighted integral relevant to $K_L \to \pi^0 e^+ e^-$.
In this respect the possibility to vary one parameter, in order to 
test the stability of the result, represents an important 
improvement with respect to the choice made in \cite{DG}.
The latter is recovered  as a special case for $b = -a =1$.

Following \cite{DG} we can write the CP-conserving 
$K_L(p) \to \pi^0(p_\pi) e^+(k_1) e^-(k_2)$ amplitude as
\beq
M\,(\,K_{L}\rightarrow \pi ^{0}\,e^{+}\,e^{-})_{\rm CPC}=   
{G_{8}\alpha^{2} B(z) G(z) \over 16 \pi^2 m_K^2} 
~ p\ccdot(k_{1}-k_{2})(p+p_\pi)_{\mu }
  \overline{u}(k_{2})\gamma ^{\mu }v(k_{1})~,
\label{eq:CPC1}
\eeq
where $z=s/m_K^2=(k_1+k_2)^2/m_K^2$ and $G(z)$ is a dimensionless function 
resulting from the loop integration. Neglecting terms which are suppressed 
by inverse powers of $m_V^2$ and eliminating $b$ by means of Eq.~
(\ref{eq:sumrule}),
we find:
\beq
G(z) =  \frac{2}{3}
\ln \left(\frac{m_{\rho }^{2}}{-s}\right) 
-\frac{1}{9}+\frac{4}{3}\left( 1+ a \right)~. 
\label{eq:CPC_G}
\eeq 
The function $B(z)$ is certainly  real for $z< 4 m_\pi^2/m_K^2$.
Above the $\pi\pi$ threshold this is no longer true; however, 
as discussed above, if $|B(z)|\gsim 1$ the local 
contribution is dominant and, to a good accuracy, $B(z)$
can be approximated by a real constant in the entire phase space. 
For this reason the product $|B(z)|\times \Im G(z)$ (where $\Im G(z)$
is obtained from (\ref{eq:CPC_G}) 
by the prescription $\ln(-m^2_\rho)=\ln m^2_\rho+i\pi$) represents to a 
good approximation the full absorptive contribution of the 
CP-conserving $K_L \to \pi^0 e^+ e^-$ amplitude.

Our result for this absorptive part is consistent with the 
one in \cite{FR89,Se93} but not with the result of \cite{DG}. 
We disagree with \cite{DG} also in the 
dispersive part, $|B(z)| \times \Re G(z)$, computed for $a =-1$, 
where our form factor is identical to the one of \cite{DG}.
The main difference between our result and the one in \cite{DG}
is that we do not find any singularity in the limit $m_e\to 0$.
The lacking of this kind of singularity was already noticed 
and discussed in \cite{FR89}. Indeed
it is possible to explicitly check that there are no possible
bremsstrahlung contributions that could cancel such a 
singular behaviour.

Integrating the amplitude (\ref{eq:CPC1}) over the phase 
space, at fixed $z$, leads to  
\beqa
&& \frac{d}{dz} B(K_L \to \pi^0 e^+ e^-)_{\rm CPC} =
\frac{ G_8^2 \alpha^4 m_K^5}{ 15\pi^7 2^{16} \Gamma_L} 
\left| B(z) G(z) \right|^2 \lambda(1,r_{\pi }^{2},z)^{5/2} \no\\
&& \qquad\qquad  = 8.1 \times 10^{-13} \times |B(z) |^{2} 
\lambda(1,r_{\pi }^{2},z)^{5/2} 
\left[ 1 + \left(\frac{3}{2\pi} \Re G(z) \right)^2 \right]~.
\eeqa
As we shall discuss in more details in Section~\ref{sect:rates},
the kinematical factor $\lambda^{5/2}$ strongly suppresses the high-$z$
contributions, strengthening the validity of the approximation 
$B(z) \simeq B(0)$. In this limit, integrating over the full 
phase space we obtain
\beqa
B(K_L \to \pi^0 e^+ e^-)_{\rm CPC} &=& 7.0 \times 10^{-14} 
\times \left| B(0) \right|^2 
\no \\
&& \times \left\{ 1 + \left[1.4 +1.4(1+a) + 
    0.4 (1+a)^2\right]\right\}~.
\label{eq:BR_CPC2}
\eeqa
The term between square brackets in (\ref{eq:BR_CPC2}) represents the 
model-dependent contribution of the dispersive amplitude: we express 
it as a function of $(1+a)$, since $(a+1) \approx 0$, 
corresponding to the choice of Ref.~\cite{DG}, 
denotes the most natural possibility. As can be noticed, 
for reasonable values of the form-factor parameters,
namely for both $a$ and $b=-1-2a$ of $\cO(1)$, 
absorptive and dispersive contributions are 
of the same order. Assuming the term between curly 
brackets in (\ref{eq:BR_CPC2}) to be smaller than 10, 
which we consider a rather conservative hypothesis, and using 
the inequality $|B(0)|^2 \leq R_B$ we can finally write 
\beq
B(K_L \to \pi^0 e^+ e^-)_{\rm CPC} < 3.5 \times 10^{-4} \times  
B(K_L \to \pi^0 \gamma\gamma)_{M_{\gamma \gamma} < 110~{\rm MeV}}~.
\label{eq:BR_CPC_f}
\eeq
Using the recent experimental input \cite{NA48_KLpgg} (see the Appendix)
\beq
B(K_L \to \pi^0 \gamma\gamma)_{M_{\gamma \gamma} < 110~{\rm MeV}} 
< 0.9 \times 10^{-8} \quad{\rm at }\quad  90\, \% {\rm C.L.}, 
\eeq
our present estimate of the maximal CP-conserving 
contribution to $K_L \to \pi^0 e^+ e^-$ reads
\beq
B(K_L \to \pi^0 e^+ e^-)_{\rm CPC} < 3 \times 10^{-12}~.
\label{eq:CPC_final}
\eeq 
We stress that Eq.(\ref{eq:CPC_final}) is a conservative upper 
bound. A reasonable estimate of $B(K_L \to \pi^0 e^+ e^-)_{\rm CPC}$, 
based on the assumption that real and absorptive contributions are 
equal, would lead to values below $10^{-12}$.

To conclude, we recall that Eq.~(\ref{eq:BR_CPC_f})
has been obtained under the assumption $|B(0)|\!~{\gsim\!}~1$.
Thus, according to Eq.~(\ref{eq:B110}), it makes sense only if 
$B(K_L \to \pi^0 \gamma\gamma)_{M_{\gamma \gamma} < 110~{\rm MeV}}   
\gsim 2 \times 10^{-9}$.
If experiments would find that 
$B(K_L \to \pi^0 \gamma\gamma)_{M_{\gamma \gamma} < 110~{\rm MeV}}$
is below this figure, other sub-leading amplitudes could 
become relevant. In particular, we have neglected the
contribution induced by $y^2$-dependent terms in $S$. The latter 
leads to a CP-conserving amplitude 
\beq
M\,(\,K_{L}\rightarrow \pi ^{0}\,e^{+}\,e^{-})_{\rm CPC}^{\rm S-type}
\approx  {G_{8}\alpha^{2} \over 16 \pi^2 m_K^2} 
\left(\frac{\partial^2 S}{ \partial y^2}\right)  
\frac{z}{12}\, p\ccdot(k_{1}-k_{2})(p+p_\pi)_{\mu }\overline{u}(k_{2})
\gamma ^{\mu }v(k_{1})~,
\label{eq:CPC_A}
\eeq
which is not only suppressed by the smallness of $\partial^2 S/\partial y^2$,
but also by extra kinematical and numerical factors (the
coefficient $z/12$ arises from the calculation with a point-like
form factor).
These contributions become relevant 
only if $B(K_L \to \pi^0 e^+ e^-)_{\rm CPC}$ is in the $10^{-13}$ range and
thus small enough compared to the CP-violating term.

\section{Indirect-CP-violating amplitude and total rate}
\label{sect:CPmix}

The CP-conserving transitions $K_{S}\rightarrow \pi ^{0} \ell^+ \ell^-$ and 
$K^{\pm }\rightarrow \pi ^{\pm } \ell^+ \ell^-$, with the 
lepton pair in a vector state, are dominated by the 
long-distance process 
$K(p)\to\pi\gamma \to \pi(p_\pi)  \ell^+(k_1) \ell^-(k_2)$ \cite{EPR}.
The decay amplitudes can in general be written as 
\beq
A\left( K_i \rightarrow \pi \ell^+ \ell^- \right) = - \frac{%
\displaystyle e^2}{\displaystyle m_K^2 (4 \pi)^2} W_i(z) (p+p_\pi)^\mu 
\bar{u}(k_2) \gamma_\mu v(k_1)~,\label{eq:CPV_S}
\eeq
where $W_i(z)$ are form factors regular at $z=0$ ($i=\pm,S$).
The latter can be decomposed as a sum of a polynomial piece plus a
non-analytic term, $W_{i}^{\pi \pi}(z)$, generated  by  
the $\pi \pi $ loop  and completely determined in terms 
of the physical $K\rightarrow 3 \pi$ amplitude \cite{DEIP}.
Expanding the polynomial term up to  $\cO(p^{6})$
we can write 
\beq
W_{i}(z)\,=\,G_{F}m_{K}^{2}\,(a_{i}\,+\,b_{i}z)\,+\,W_{i}^{\pi \pi
}(z)\;,
\label{eq:Wp6}
\eeq
where the real parameters $a_{i}$ and $b_{i}$ encode local 
contributions starting at $\cO(p^{4})$  
and $\mathcal{O(}p^{6})$, respectively. 
High-precision data on $K^{+}\rightarrow \pi ^{+}e^{+}e^{-}$ by BNL-E865
\cite{kpeeE865} have been successfully fitted using Eq.~(\ref{eq:Wp6})
and lead to
\beq
a_{+}\,=-0.587\pm 0.010\qquad \,b_{+}=-0.655\pm 0.044~.  
\label{eq:ab+}
\eeq
Unfortunately, chiral symmetry alone does not help to 
determine the unknown couplings $a_{S}$ and $b_S$ 
in terms of $a_{+}$ and $b_{+}$. On the other hand, 
the non-analytic term $W_{S}^{\pi \pi}(z)$ is known to 
be very small, due to the $\Delta I=3/2$ suppression 
of the CP-conserving $K_S \to 3\pi$ amplitude. As a consequence, 
the $K_S \rightarrow \pi^0 e^+ e^-$ rate turns out to 
be dominated by local contributions 
\beqa
B(K_S \rightarrow \pi^0 e^+ e^-)  &=&  \left[ 0.01 - 
0.76 a_S - 0.21 b_S
+ 46.5 a_S^2 + 12.9 a_S b_S + 1.44 b_S^2 \right]\times 10^{-10} \no\\
  & \approx &    5.2 \times 10^{-9} \times a_S^2~,
\label{eq:BRKS}
\eeqa
where the second line follows from the assumption 
$b_S/a_S =m^2_K/m^2_\rho$, 
motivated by VMD.

The first experimental evidence of the $K_S \rightarrow \pi^0 e^+ e^-$ transition 
has been announced very recently by the NA48/1 Collaboration at CERN.
The observation of 7 events in a clean signal region (with 0.15 expected 
background events) leads to the preliminary result~\cite{NA48KS}:
\beq
B(K_S \rightarrow \pi^0 e^+ e^-)_{m_{ee} > 165~{\rm MeV}} = \left( 3.0^{+1.5}_{-1.2} \pm 0.2 \right) \times 
10^{-9}~,
\eeq
which implies 
\beq
|a_S|= 1.08^{+0.26}_{-0.21}~.
\label{eq:as}
\eeq
This result is in good agreement with the naive chiral 
counting expectation $a_S = \cO(1)$~\cite{DEIP}
and, within this framework, it is quite compatible with the 
ranges expected from large-$N_C$~\cite{Bruno} and resonance-saturation 
arguments~\cite{EKM93}. Interestingly, the central value of the NA48/1 result 
is also in excellent agreement with a very old prediction 
by Sehgal, based only on VMD and $\Delta I=1/2$ rule \cite{Sehgal_old}.

Taking into account the interference between  direct- and indirect-CP-violating components,  
the full CP-violating contribution to $K_L \rightarrow \pi^0 e^+ e^-$ can 
be written as 
\beq
B(K_L \rightarrow \pi^0 e^+ e^-)_{\rm CPV} \, = \, 10^{-12} \times  \left[ C_{\rm mix} 
\pm C_{\rm int}  \left( \frac{\displaystyle \Im \lambda_t}{\displaystyle 10^{-4}}
\right) + C_{\rm dir}  \left( \frac{\displaystyle \Im \lambda_t}{\displaystyle 10^{-4}}
\right)^2 \right]~, \label{eq:cpvt}
\eeq
where the $\pm$ depends on the relative sign between short- and 
long-distance amplitudes, and 
\beqa
 C_{\rm mix} &=& 10^{12}\,
|\epsilon|^2  \frac{\tau(K_L)}{\tau(K_S)} B(K_S \rightarrow \pi^0 e^+ e^-) = 
(15.7 \pm 0.3) |a_S|^2~, \no \\
 C_{\rm dir} &=& 
10^4 \frac{\tau(K_L)}{\tau(K^+)} \frac{B(K^+_{e3})}{|V_{us}|^2}
( y_{7A}^2 + y_{7V}^2 ) = 2.4 \pm 0.2 ~, \no\\
C_{\rm int} &=& 2 \sin\phi_\epsilon \sqrt{ C_{\rm mix} C_{\rm dir} }
\frac{ y_{7V} }{ \sqrt{ y_{7A}^2 + y_{7V}^2 } } {\cal F} = (6.2\pm0.3) |a_S|~,
\label{eq:cmix}
\eeqa
with $\phi_\epsilon = 43.7^\circ$. The
numerical expressions for $C_{\rm mix}$ and $C_{\rm int}$
in terms of $|a_S|$ are computed assuming $b_S/a_S=m^2_K/m^2_\rho$
(or the same form factor for  direct- and indirect-CP-violating components)
and considering only the quadratic terms in (\ref{eq:BRKS});
the quoted error reflects the impact of the linear terms.
On the other hand, the l.h.s equations in (\ref{eq:cmix})
are valid independently of any assumption on $W_S(z)$, 
whose possible difference from the  
short-distance form factor $f_+(z)$ 
is taken into account by the factor
\beq
{\cal F} = \frac{ \int dz  \lambda^{3/2}(1,z,r_{\pi}^2)  
\left[ {\Re}\left[ W^*_S(z) f_+(z) \right]
     - {\Im}\left[ W^*_S(z) f_+(z) \right]\right]  }{
\left[ \int dz  \lambda^{3/2}(1,z,r_{\pi}^2) | W_S(z) |^2 \right]^{1/2} \times 
\left[ \int dz  \lambda^{3/2}(1,z,r_{\pi}^2) | f_+(z) |^2  \right]^{1/2}  }~,
\eeq
where $r_\pi = m_\pi/m_K$ and $\lambda(a,b,c)=a^2+b^2+c^2-2(ab+ac+bc)$.
We also remark that in evaluating $B(K_S\to\pi^0e^+e^-)$,
$\alpha=1/137$ has been used for the electromagnetic coupling.
For a given branching ratio, a different choice of $\alpha$ implies,
effectively, a different normalization of $|a_S|$ without changing
the estimate of $C_{\rm mix}$.

Given the recent result in (\ref{eq:as}), we can 
conclude that, within the SM, the indirect-CP-violating amplitude 
is the largest component of $B(K_L \rightarrow \pi^0 e^+ e^-)$
and the full CP-violating contribution is 
well above the CP-conserving one.

\subsubsection*{Sign of the interference term}

As can be seen from (\ref{eq:cpvt}), the 
relative sign between short- and long-distance contributions 
is a crucial ingredient for estimating the sensitivity of 
future high-statistics experiments to perform precision 
SM tests in this mode. A prediction of this sign 
requires a better understanding of the dynamical origin 
of the local couplings $a_i$ and $b_i$. 
To this purpose, we first note that 
the experimental determination of the ratio $b_{+}/a_{+}$
does not strictly follow naive dimensional analysis,
which would predict $b_{+}/a_{+} = \cO[m^2_K/(4\pi F_\pi)^2] \sim 0.2$.
Such a large ratio between $\cO(p^4)$ and 
$\cO(p^6)$ counterterms naturally points to the presence 
of large VMD contributions in these channels.
Under the rather general hypothesis 
that the $b_{i}$ terms in Eq.~(\ref{eq:Wp6}) 
are entirely generated by the expansion of 
a vector-meson propagator, we can re-write the polynomial contribution
to $W_i(s)$ as
\beq  
G_{F}m_{K}^{2}\left( \frac{a_{i}^{\rm VMD}}{1-z m_{K}^{2}/m_{V}^{2} } 
+ a_{i}^{\rm nVMD} \right) 
\approx  G_{F}m_{K}^{2} \left[  \left(a_{i}^{\rm VMD} + 
   a_{i}^{\rm nVMD} \right)
+ a_{i}^{\rm VMD} \frac{m_{K}^{2}}{m_{V}^{2}} z \right]~,
\label{eq:poleexp}
\eeq
where $a_{i}^{\rm nVMD}$ denotes $z$-independent 
non-VMD contributions.\footnote{~A similar 
generic decomposition between VMD and non-VMD contributions has been 
successfully tested in the $K_L \rightarrow \pi^+ \pi^- \gamma$ mode 
\cite{KTeV-kppg,gao00}.}
A comparison with the experimental results in Eq.~(\ref{eq:ab+}) then 
leads to
\beq
a_+^{\rm VMD}=\frac{m_{V}^{2}}{m_{K}^{2}}b_+^{\rm exp}=-1.6 \pm 0.1~, \qquad 
a_+^{\rm nVMD} = a_+^{\rm exp} - a_+^{\rm VMD} = 1.0 \pm 0.1~.
\label{eq:a_+VMD}
\eeq
The large value of $a_+^{\rm nVMD}$ can be justified, in the charged channel, 
by the presence of sizable pion-loop contributions.
On the contrary, one would expect a pure VMD contribution 
in the $K_S$ mode, where the pion term is negligible. 
We thus expect  $a_S^{\rm nVMD}=0$ and $a_S^{\rm VMD}=a_S$, 
an assumption that justifies the use of the same form factor 
for direct- and indirect-CP-violating components
adopted to derive (\ref{eq:cpvt}).

To make a theoretical prediction for  $a_S$ and, particularly, 
to determine its sign, we need to take an additional step. 
The contributions to the amplitude (\ref{eq:CPV_S})
generated by the leading short-distance operator in Eq.~(\ref{eq:heff}), 
namely $Q_{7V}$, satisfy the $\Delta I=1/2$ isospin relation
\beq
(a_S)_{\langle Q_{7V} \rangle} =-(a_+)_{\langle Q_{7V} \rangle }~.
\eeq
If this relation is obeyed by the full VMD amplitude, 
we should expect
\beq
(a_S^{\rm VMD})_{\langle Q_{7V} \rangle} = -a^{\rm VMD}_+ = 1.6 \pm 0.1~.
\label{eq:as_pred}
\eeq
Given the various assumptions behind this prediction,
the agreement with the experimental result in (\ref{eq:as})
is rather encouraging.\footnote{~The $\Delta I=1/2$ relation between $\cO(p^4)$ local 
counterterms of $K^+ \rightarrow \pi^+ e^+ e^-$ and $K_S \rightarrow \pi^0 e^+ e^-$
amplitudes had already been adopted in Ref.~\cite{EPR} and later on 
also in Ref.~\cite{EKM93}. However, in these works 
no distinction is made between VMD and non-VMD contributions 
to $a_+$, which turns out to be a fundamental ingredient to 
obtain a phenomenologically acceptable prediction for 
$a_S$.}
This brings us to the likely hypothesis that the VMD part 
of the $W_i(z)$ is dominated by the 
CP-conserving contribution induced by $Q_{7V}$,
in the limit of a low matching 
scale ($m_K \lsim \mu \lsim m_V$).\footnote{~Note that, consistently
with this hypothesis, the contributions generated 
by $Q_{7V}$ satisfy the
relation $(b_+/a_+)_{\langle Q_{7V} \rangle } = (b_S/a_S)_{\langle Q_{7V} \rangle} 
= m_{K}^{2}/m_{V}^{2}$, 
which follows from the presence of the vector 
form factor $f_+(z)$ in Eq.~(\ref{eq:SD_ampl}). }
The form factors computed under this assumption
from the short-distance Hamiltonian cannot be trusted in detail, 
since  $z_{7V}(\mu)$ exhibits a strong scale dependence 
(that should be matched by the matrix-elements 
of four-quark operators)
and we would like to extrapolate it beyond the validity region of
perturbation theory. However, the VMD structure of the form factors
ensures we don't 
need to push this scale too low. Moreover, we shall employ this procedure 
only to fix the relative sign between direct- and indirect-CP-violating 
contributions. Using  the $ Q_{7V}$ part of the short-distance 
Hamiltonian to determine the full $K_L \rightarrow \pi^0 e^+ e^-$
 vector form factor, we find 
\beq
W_L(z) = G_F m_K^2 \frac{4\pi  y_{7V}(\mu) }{\sqrt{2}\alpha} f_+(z)
\left[   |\epsilon| e^{i\phi_\epsilon}
  V_{us} \frac{z_{7V}(\mu) }{ y_{7V}(\mu)} - i  \Im \lambda_t \right]~.
\eeq
Since the ratio $z_{7V}(\mu)/y_{7V}(\mu)$ is negative for $\mu < 1$~GeV \cite{BLMM,BBL}, 
we conclude that the most likely possibility is a {\em positive} interference 
between direct- and indirect-CP-violating components. 

The prediction of a positive interference between direct- and 
indirect-CP-violating 
components of $K_L \rightarrow \pi^0 e^+ e^-$ had already been reached in 
Ref.~\cite{GW}.
This conclusion is reinforced by the observation that the 
perturbative value of  $a_S$, computed from  $\langle Q_{7V} (\mu) \rangle$,
is smaller than the experimental result in (\ref{eq:as}) for $\mu=1$~GeV, 
but it grows, i.e.~it goes in the right direction, for smaller values of 
$\mu$.

The consistency of this prediction can be further confirmed by the 
following complementary reasoning: if we trust the  $\Delta I=1/2$ + VMD 
argument leading to Eq.~(\ref{eq:as_pred}), we have fully established the 
sign of $a_S$ within chiral perturbation theory. This sign follows from 
the sign of $a_+$ which, in turn, is determined experimentally 
by the interference between local and non-local terms in $(\ref{eq:Wp6})$. 
We have thus established the sign of $a_S$ in terms of the sign of 
$G_8$, the overall coupling of the $(8_L,1_R)$ non-leptonic
weak chiral Lagrangian (see e.g.~Ref.~\cite{Ecker}). This implies that 
the $K_L \rightarrow \pi^0 e^+ e^-$ vector form factor 
can be written as
\beq
W_L(z) = G_F m_K^2 f_+(z) \left[ a_S  |\epsilon| e^{i\phi_\epsilon}{\rm sgn}(G_8)
- i \frac{4\pi  y_{7V}(\mu) }{\sqrt{2}\alpha}  \Im \lambda_t \right]~,
\label{eq:WL2}
\eeq
with a positive $a_S$. The sign of $G_8$
cannot be determined in a model-independent way; however, 
it can be predicted by the partonic Hamiltonian in 
(\ref{eq:heff}) employing naive factorization. By doing so 
we find ${\rm sgn}(G_8) <0$ confirming the {\em positive} interference 
between direct- and indirect-CP-violating components.

\subsubsection*{Total rate and present bounds on $\Im \lambda_t$}

\begin{figure}[t]
\begin{center}
\leavevmode
\epsfxsize=12.0 cm
\epsfbox{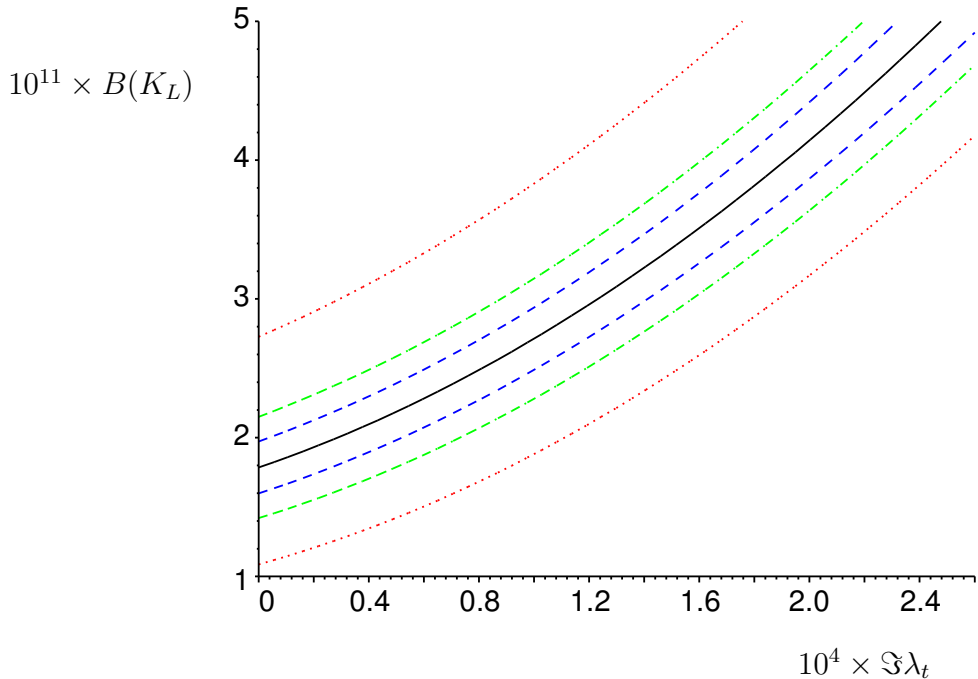}
\end{center}
\vskip -8.6 cm
\hskip  0.2 cm $10^{11}\times B(K_L)$
\vskip  7.2 cm
\hskip  10.7 cm $10^{4} \times \Im \lambda_t$
\vskip  0.4 cm
\caption{SM Prediction for $B(K_L \rightarrow \pi^0 e^+ e^-)$
as a function of $\Im \lambda_t$, neglecting CPC contributions and 
assuming a positive interference between direct- and indirect-CP-violating 
components (see text). The three curves correspond to a central value 
$a_S= 1.08$ and no error (central full line); $5\%$ error  (dashed blue lines);
$10\%$ error  (dashed green lines); present error (red dotted lines). }
\label{fig:KL_rate}
\end{figure}

Employing the positive sign in (\ref{eq:cpvt}) we finally arrive 
at the prediction
\beq
B(K_L \rightarrow \pi^0 e^+ e^-)_{\rm SM} \, = 
\left( 3.2^{+1.2}_{-0.8} \right) \times 10^{-11}
\label{eq:BKL_SM}
\eeq
where the error is completely dominated by the uncertainty in 
$B(K_S \rightarrow \pi^0 e^+ e^-)$ and, in view of (\ref{eq:CPC_final}), 
we have neglected the CP-conserving term. 
The possible improvements of this prediction 
with better data on the $K_S$ mode,
and the related sensitivity to $\Im \lambda_t$, 
are illustrated in Fig.~\ref{fig:KL_rate}. 

At present, Eq.~(\ref{eq:BKL_SM}) has to be compared 
with the new preliminary upper limit by KTeV
\beq
B(K_L \rightarrow \pi^0 e^+ e^-) < 2.8 \times 10^{-10}
\quad{\rm at }\quad  90\, \% {\rm C.L.}, 
\eeq
based on the full 1997+1999 data sample \cite{KTeV_p0ee}.
Although we are still far from the possibility of precision SM tests
in this mode, the combination of the preliminary 
results by KTeV and NA48/1 allows us to derive significant bounds 
on realistic non-standard scenarios (see e.g.~Ref.~\cite{BCIRS}).
Instead of discussing any specific model in detail, 
we can simply express these bounds as the limits on $\Im \lambda_t$
dictated by  $K_L \rightarrow \pi^0 e^+ e^-$.
In particular, we find 
\beq
- 1.3 \times 10^{-3} < \Im \lambda_t < 1.0 \times 10^{-3} 
\quad{\rm at }\quad  90\, \% {\rm C.L.}, 
\eeq
which would reduce to $|\Im \lambda_t| < 1.3 \times 10^{-3}$ in the absence 
of any assumption about the interference term in (\ref{eq:cpvt}).
Together with the complementary and comparable 
constraints on $\lambda_t$ derived by 
$K^+ \to \pi^+ \nu \bar{\nu}$ \cite{E787}, these limits 
represent the most precise information presently available 
about the short-distance structure of the FCNC $s\to d$ 
amplitude.

\section{Dalitz-plot analysis}
\label{sect:rates}

The most serious problem in the extraction of the  $K_L \to \pi^0 e^+ e^-$
amplitude from a time-independent 
rate measurement is the large irreducible background generated
by the process $K_L \to \gamma \gamma e^+ e^-$ \cite{Greenlee}.
Imposing the cut $|M_{\gamma \gamma}-m_{\pi^0}|<5$~MeV
on the two-photon invariant mass spectrum 
of $K_L \to \gamma \gamma e^+ e^-$, the latter 
turns out to have a branching ratio $\sim 3 \times 10^{-8}$,
more than $10^3$ times larger than the signal.
As discussed by Greenlee \cite{Greenlee}, 
employing additional cuts on various 
kinematical variables it is possible to 
reduce this background 
down to the  $10^{-10}$ level  
\cite{Greenlee,KTeV_p0ee}, but it is hard 
to reduce it below this figure without 
drastic reductions of the signal efficiency. 
We stress, however, that this does not 
imply that the signal is unmeasurable 
in a high-statistic experiment, 
where the physical background can be measured 
and modelled with high accuracy \cite{muon}.

An important point to notice is that 
the kinematical analysis necessary to suppress and control 
the Greenlee background provides also a powerful tool to 
discriminate the CP-conserving component of the 
$K_L \rightarrow \pi^0 e^+ e^-$ rate. As discussed 
in Section~\ref{sect:CPC}, theoretical arguments 
suggest that the CP-conserving component 
of $K_L \rightarrow \pi^0 e^+ e^-$ is very small;
however, it is clearly desirable to cross-check this 
statement {\em a posteriori}, in a model-independent way,
using  $K_L \rightarrow \pi^0 e^+ e^-$ data.

\begin{figure}[p]
\begin{minipage}[t]{77mm}
  \begin{center}
  \leavevmode
  \epsfysize=7.0 cm
  \epsfxsize=7.0 cm
  \epsfbox{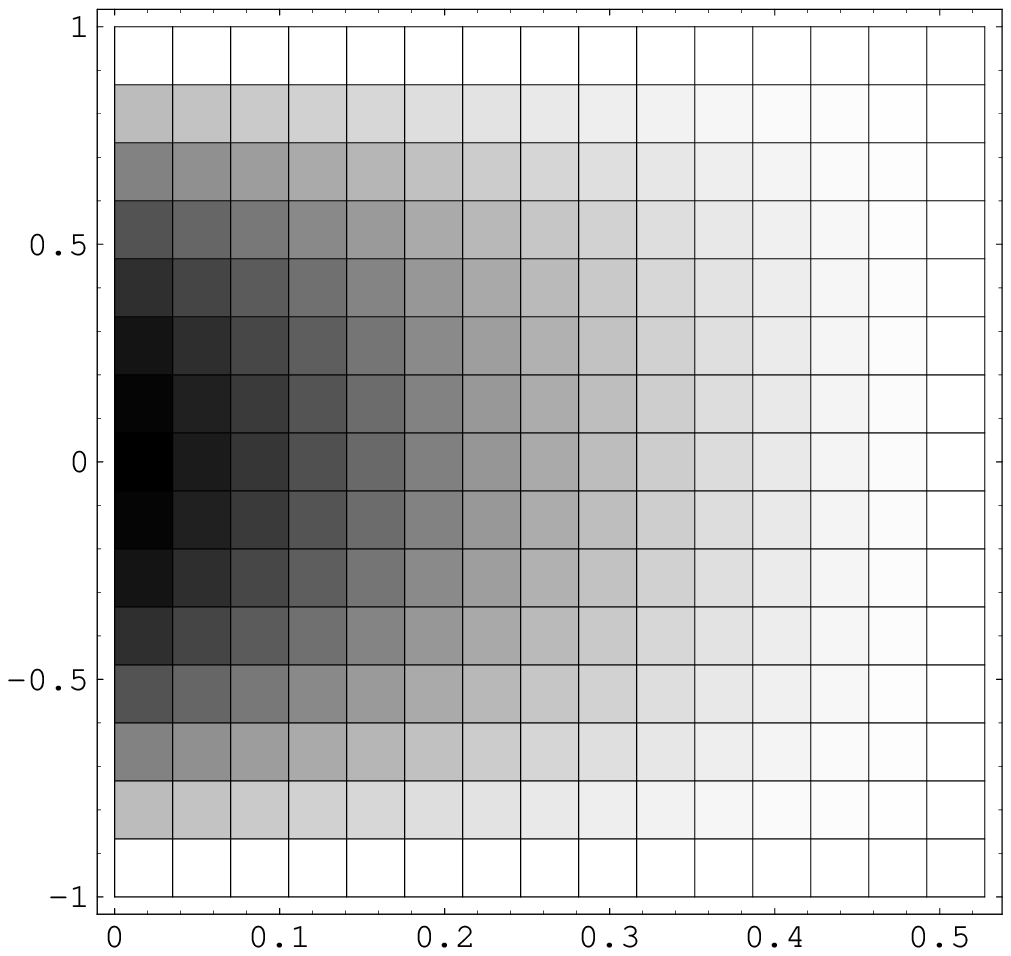}
  \end{center}
\end{minipage}
\hspace{\fill}
\begin{minipage}[t]{77mm}
  \begin{center}
  \leavevmode
  \epsfysize=7.0 cm
  \epsfxsize=7.0 cm
  \epsfbox{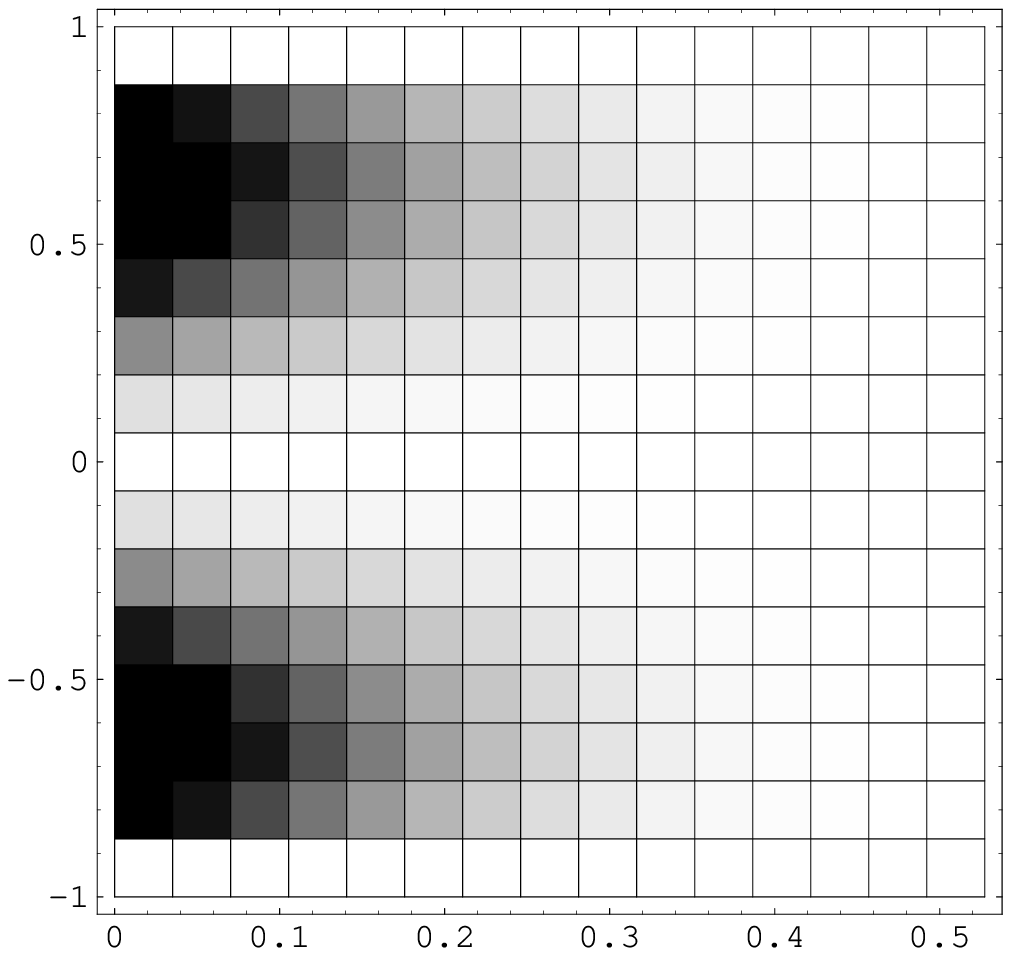}
  \end{center}
  \end{minipage}
  \label{fig:dalitz}
  \caption{Dalitz plot distributions for CP-violating (left) and 
           CP-conserving (right) contributions 
           to $K_L \rightarrow \pi^0 e^+ e^-$ ($\tilde y \to$ vertical axis,
           $z \to $ horizontal axis).}
\vskip  2.5 cm 
\hskip  1.5 cm $\displaystyle{ \int \frac{d \Gamma }{ d {\tilde y} }}$
\begin{center}
\leavevmode
\vskip -4.0 cm
\hskip  0.5 cm  
\epsfysize=8.5 cm
\epsfxsize=10 cm
\epsfbox{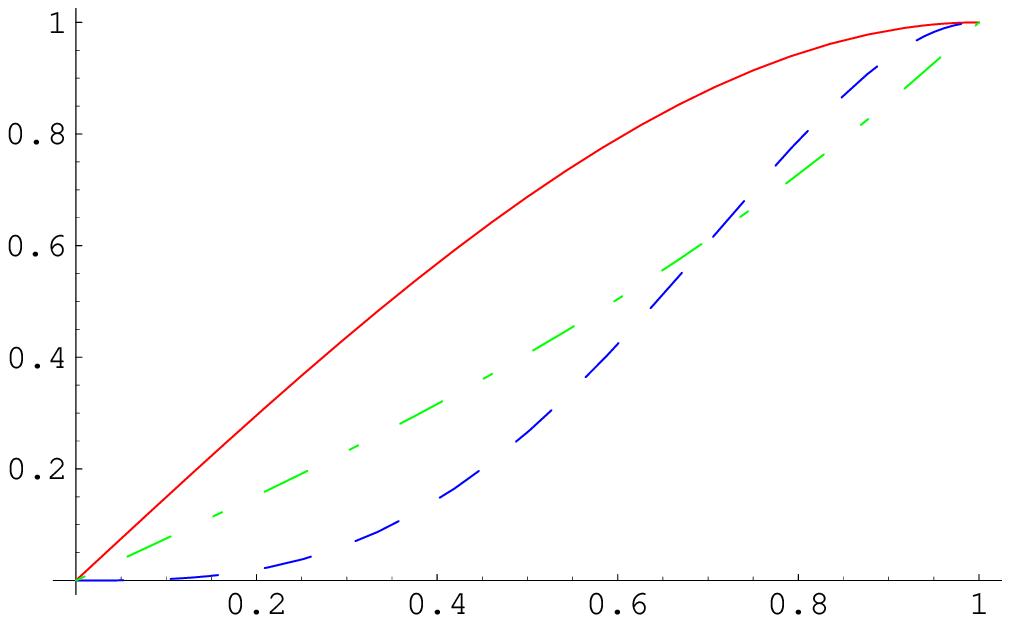}
\end{center}
\vskip -0.4 cm
\hskip  12 cm $|\tilde y|_{\rm max}$
\vskip  0.5 cm 
\caption{CP-violating (full line) and CP-conserving (dashed line)
contributions to the $K_L \rightarrow \pi^0 e^+ e^-$ rate 
(with arbitrary normalization) as a function of $|\tilde y|_{\rm max}$. 
The dot-dashed line denotes Greenlee background.}
\label{fig:y2cutint}
\end{figure}

As discussed by Greenlee \cite{Greenlee}, the most convenient 
kinematical variables to describe the decay distribution of 
$K_L \rightarrow \pi^0 e^+ e^- $ are 
\beq
z = \frac{ (k_{1}+k_{2})^{2} }{ m_{K}^{2}}~, \qquad 
{\tilde y} = \frac{ 2~ p\ccdot (k_{2}-k_{1}) }{ 
m_K^{2} \lambda^{1/2}(1,r_{\pi }^{2},z) }~,
\end{equation}
whose uncorrelated boundaries (in the limit $m_e/m_K \to 0$) are given by 
\beq
 0 \leq z \leq (1-r_{\pi})^2 \quad \quad -1< {\tilde y} <1~.
\eeq
In terms of these variables, CP-conserving and CP-violating 
distributions assume the following simple factorized structure:
\beqa
\frac{d^2 \Gamma(K_L \rightarrow \pi^0 e^+ e^-)_{\rm CPV} }
{ d z ~d {\tilde y}} & \propto &
(1- {\tilde y}^2) \lambda^{3/2}(1,z,r_{\pi}^2)  \left|W(z)\right|^2~, \\
\frac{d^2 \Gamma(K_L \rightarrow \pi^0 e^+ e^-)_{\rm CPC} }
{ d z ~d {\tilde y}} & \propto &
{\tilde y}^2 (1- {\tilde y}^2) \lambda ^{5/2}(1,z,r_{\pi}^2) 
\left| B(z) G(z) \right|^2~.
\eeqa
In Fig.~\ref{fig:dalitz} we plot the two distributions, 
as obtained by setting $W(z) = 1+ z m_{K}^{2}/m_{V}^{2}$ 
and approximating $B(z)G(z)$ to a constant:
the ${\tilde y}$ dependence clearly provides a powerful 
tool to discriminate the two terms. 

To better quantify to which extent the
two distributions can be distinguished, 
in Fig.~\ref{fig:y2cutint} we show the result 
of a cut on $\tilde y$ ($|\tilde y| \leq |\tilde y|_{\rm max}$)
on the integrated rate. The three curves correspond to 
the three non-interfering contributions of 
CP-conserving and CP-violating amplitudes, and
Greenlee background (as obtained with a constant 
$K_L \rightarrow \gamma^* \gamma^*$ form factor), 
all normalized to one. 
As can be noted, setting the cut $|\tilde y|_{\rm max}=0.5$ the 
CP-violating rate is reduced only by $30\%$, while 
the CP-conserving one drops almost by a factor of 4.   
The cut on $|\tilde y|_{\rm max}$ in not particularly efficient 
against the Greenlee background, but also in this case it improves 
the signal/background ratio.

\section{Time-dependent interferences}
\label{sect:time}

\begin{figure}[t]
\begin{minipage}[t]{79mm}
  \begin{center}
  \leavevmode
  \epsfxsize=7.8 cm
  \epsfbox{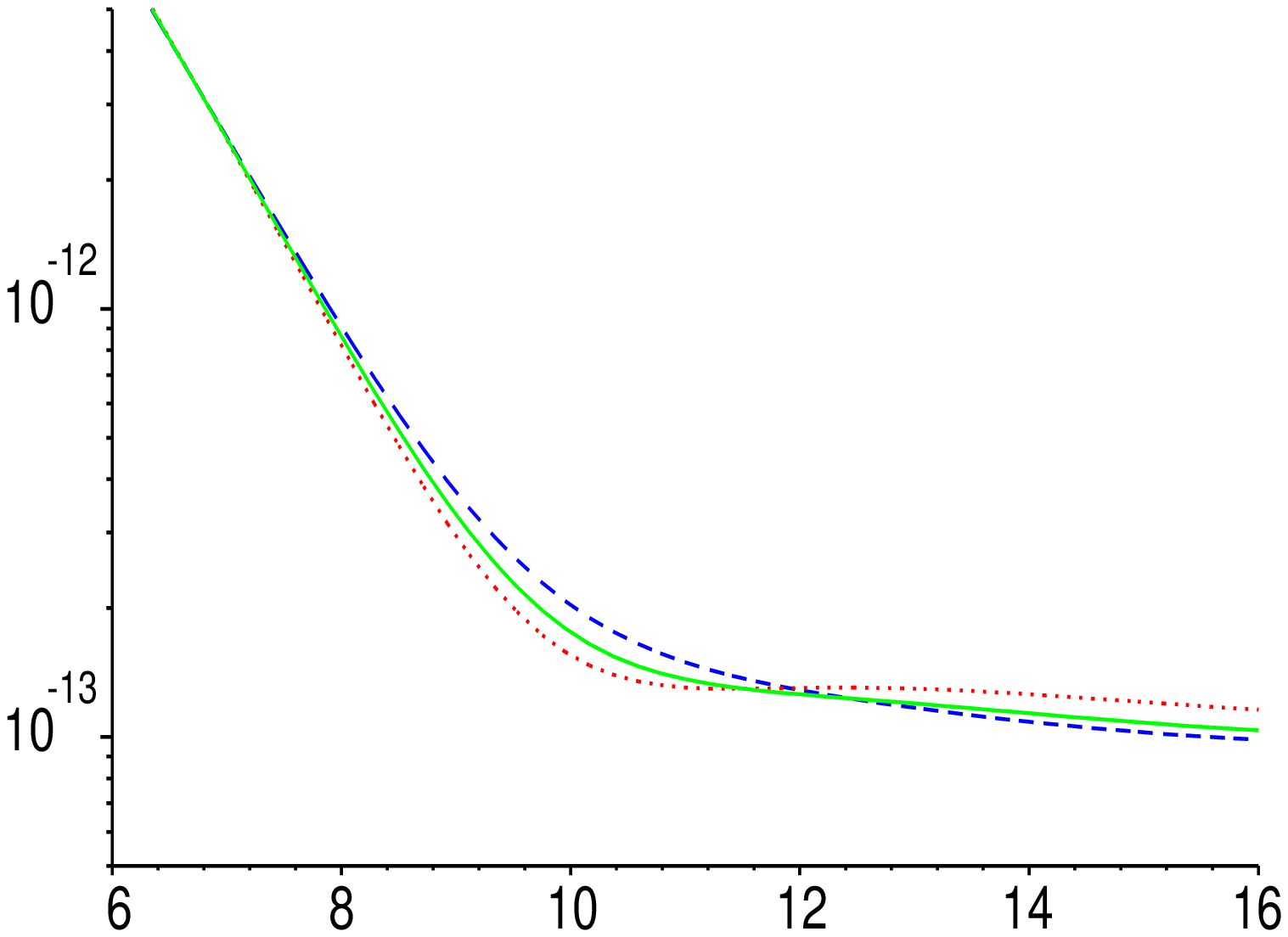}
\vskip  -6.0 cm
\hskip  -2.8 cm $ I_0(t) $
\vskip   5.0 cm
\hskip   5.8 cm $ {t/\tau_S}$
\vskip   0.4 cm
  \end{center}
\end{minipage}
\hspace{\fill}
\begin{minipage}[t]{79mm}
  \begin{center}
  \leavevmode
  \epsfxsize=7.8 cm
  \epsfbox{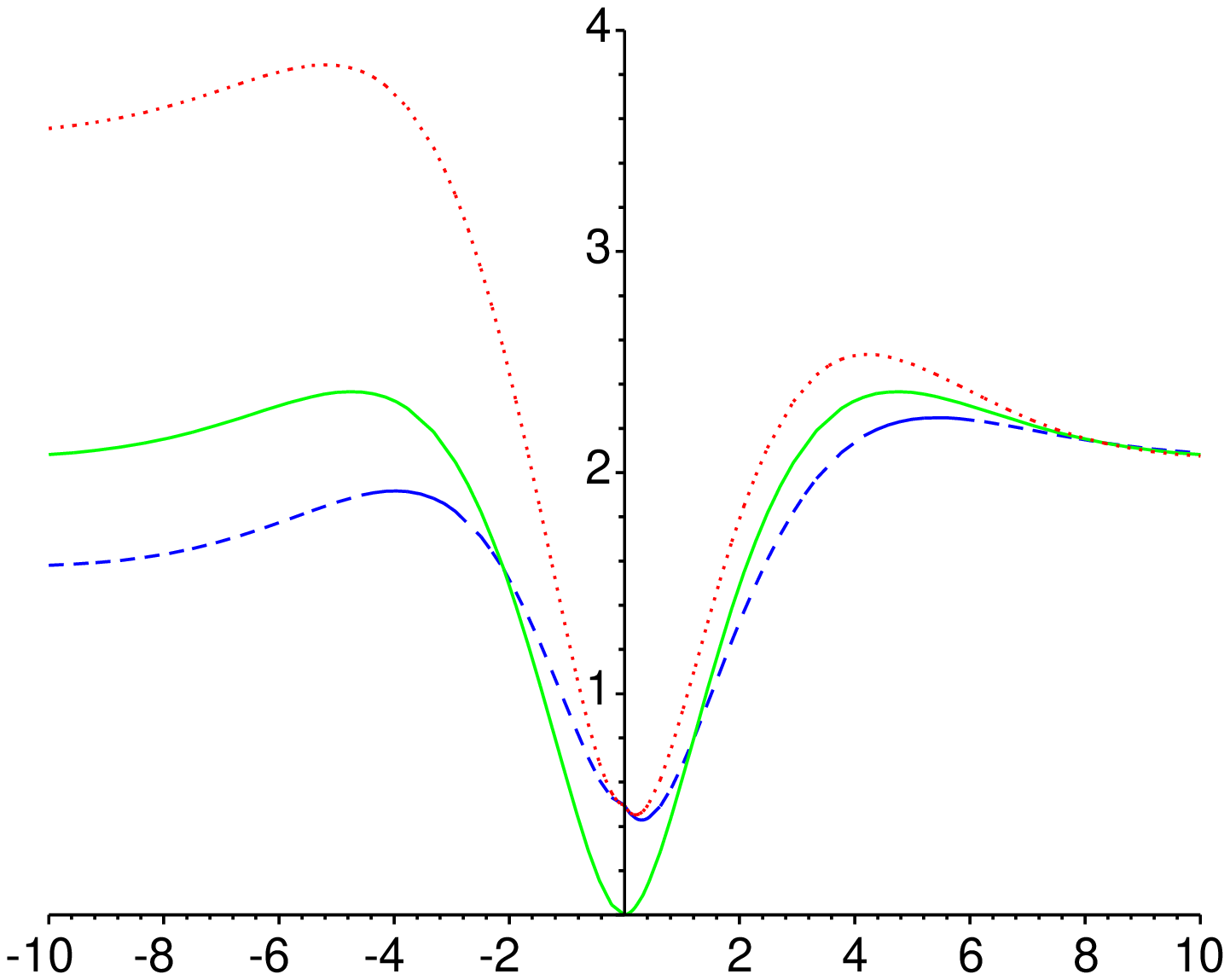}
\vskip  -6.0 cm
\hskip   3.5 cm $ 10^{14} \times I_\Phi (t) $
\vskip   5.0 cm
\hskip   0.4 cm $ {t/\tau_S}$
\vskip   0.4 cm
  \end{center}
  \end{minipage}
\caption{ Time dependent distributions 
 for ${\rm Im}(V_{ts}^*V_{td})=0$ (green full), 
$+ 1.3 \times 10^{-4}$ (red dotted) and $- 1.3 \times 10^{-4}$ (blue dashed).
Left: probability distribution for a $|K^0\rangle$ state at $t=0$ to decay
into $| \pi^0(\gamma \gamma) e^+e^-\rangle$, Greenlee background included.
Right:  Probability distribution for $\Phi \to K_L , K_S 
\to \pi^+\pi^-,\pi^0e^+e^-$ events at a $\Phi$ factory (see text).}
\label{fig:interf}
\end{figure}

Complementary information on the direct CP-violating 
component of the $K_L\to\pi^0 e^+e^-$ amplitude 
can be obtained by studying the time evolution of the 
$K_{L,S} \to \pi^0 e^+e^-$ decay \cite{DG,timeevol}.
Although challenging from the experimental point of 
view, this method has two intrinsic advantages: 
i) the interference between $K_S$ and $K_L$ amplitudes
is only due to the CP-violating part of the latter;
ii) the process $K_S \to \gamma \gamma e^+e^- $ is very 
suppressed with respect to $K_S \to \pi^0 e^+e^-$
[$B(K_S \to \gamma \gamma e^+e^-) \sim {\rm few} \times 10^{-12}$],
so the background due to the $| \gamma \gamma e^+e^- \rangle$
final state is almost negligible 
at small times ($t \ll \tau_L$). 
As representative examples of time-dependent observables, 
we shall discuss in more detail two specific cases: 
i) the time evolution of an initial $|K^0\rangle$ state, representative of
a possible fixed-target experimental set up; 
ii) the time evolution of the $K_S K_L$ coherent state produced 
at a  $\Phi$ factory \cite{D'Ambrosio:1994wx}.

In the case of a pure $|K^0\rangle$ beam at $t=0$, with $N_0$ particles, 
the number of decays into the final state 
$|\pi^0 e^+e^- \rangle$  or in the background channel 
$|e^+e^- \gamma \gamma \rangle$ (with $M_{\gamma \gamma} \sim m_{\pi^0}$),
as a function of the proper time $t$, 
can be written as
\begin{eqnarray}
I_0(t) = \tau_S \frac{d N}{ dt} &=& \tau_S \frac{N_0}{2} 
\left\{ \left| {\cal A}_S \right|^2 e^{-t/\tau_S}
+2 {\rm Re}\left[{\cal A}_L^{\rm CPV} {\cal A}_S^* e^{-i(m_L-m_S)t} 
  \right] 
e^{-t(\tau_L+\tau_S)/2\tau_L\tau_S} \frac{}{} \right. \nonumber\\ 
&& \qquad\qquad \left. 
+ \left[ \left| {\cal A}_L^{\rm CPV} \right|^2 + 
\left| {\cal A}_L^{\rm CPC} \right|^2 + \left| {\cal A}_L^{\rm bkg.} 
   \right|^2 \right]  
e^{-t/\tau_L } \right\}~,
\end{eqnarray}
where the decay amplitudes ${\cal A}_{S}$ and ${\cal A}_L^{\rm CPV}$ 
are those in Eqs.~(\ref{eq:Wp6}) and (\ref{eq:WL2}), 
and an overall phase-space integral is understood.
The three curves in Fig.~\ref{fig:interf} have been obtained for $N_0=1$,
assuming 
$\tau_S\, |{\cal A}_S |^2 = B(K_S\to\pi^0 e^+e^-)  = 6 \times 10^{-9}$ and 
$\tau_L\, |{\cal A}_L^{\rm bkg.}|^2 = B(K_L\to\gamma\gamma e^+e^-)_{\rm cuts} 
= 10^{-10}$, and employing the 
following  three values of ${\rm Im}(V_{ts}^*V_{td})$:
$0$,  $\pm 1.3 \times 10^{-4}$. 
As can be seen, the interference term is 
quite sensitive to the value of the direct CP-violating amplitude.
On a purely statistical level, in this example one could reach a $\approx 15\%$
error on ${\rm Im}(V_{ts}^*V_{td})$ 
with an initial flux of $10^{15}/\epsilon^\prime_{\pi^0ee}$
$K^0$, where $\epsilon^\prime_{\pi^0ee}$ denotes the 
efficiency for decays occurring within the first 15 $K_S$
decay lengths.

At $\Phi$ factories  we can take advantage of the quantum properties
of the $K_S K_L$ state produced by the $\Phi$ decay:  
\beq
{\cal A}\left[ \phi \rightarrow f_1(t_1),f_2(t_2) \right] = 
 \frac{1}{\sqrt{2}} \left[ {\cal A}_S(f_1,t_1) {\cal A}_L(f_2,t_2) - 
 {\cal A}_S(f_2,t_2) {\cal A}_L(f_1,t_1) \right]~,
\eeq
where $f_i$ and $t_i$ denote respectively final states and decay times.
Choosing $ f_1= | \pi ^+ \pi ^- \rangle$,  $ f_2=| \pi^0 e^+ e^-\rangle$ 
and integrating over $t_1+t_2$ \cite{D'Ambrosio:1994wx}, 
we obtain for $t=t_1-t_2$ 
\begin{eqnarray}
I_\Phi (t) &=& N_0 \frac{  B(K_S \rightarrow \pi ^+ \pi ^-)B(K_S \rightarrow \pi ^0 e^+ e^-) }{ 
\left( 1+ \tau_S/\tau_L \right)  } e^{-|t|(\tau_L+\tau_S)/2\tau_L\tau_S}  \nonumber \\
&& \times \left\{ |\eta _{+-}|^2    e^{t(\tau_L-\tau_S)/2\tau_L\tau_S }  +  \frac{ \tau_S}{ \tau_L}
\frac{ B(K_L\rightarrow \pi ^0 e^+ e^-) }{
       B(K_S \rightarrow \pi ^0 e^+ e^-) } e^{-t(\tau_L-\tau_S)/2\tau_L\tau_S } \right. \nonumber\\
&& \left.   -  \frac{ 2 \tau_S}{ B(K_S\rightarrow  \pi ^0 e^+ e^-)}
 {\rm Re}\left[  \eta _{+-}^* {\cal A}_S^* {\cal A}_L^{\rm CPV}  e^{i(m_L-m_S)t}  \right] \right\} ~.
\label{eq:I_phi}
\end{eqnarray}
Also in this case we plot in Fig.~\ref{fig:interf}
three curves corresponding to  ${\rm Im}(V_{ts}^*V_{td})=0$, and 
$\pm 1.3 \times 10^{-4}$, obtained for 
$\tau_S\, |{\cal A}_S |^2 = B(K_S\to\pi^0 e^+e^-)  = 6 \times 10^{-9}$ and $N_0=1$
but, for simplicity, here we ignore the Greenlee background. 
The direct-CP-violating effect appears to be remarkably clear, 
but the statistics necessary to measure it is probably beyond 
the near-future possibility of existing facilities.

\section{Conclusions}

Motivated by recent experimental results from NA48 on 
$K_L\to\pi^0\gamma\gamma$ and $K_S\to\pi^0 e^+ e^-$, which has been
observed for the first time, we have presented a re-analysis of the
decay mode $K_L\to\pi^0 e^+ e^-$. 

The results on $K_L\to\pi^0\gamma\gamma$ together with a more general
treatment of the $K_L\to\pi^0\gamma^*\gamma^*$ form factor have led
to an updated limit on the CP conserving contribution:
$B(K_L\to\pi^0 e^+ e^-)_{\rm CPC} < 3 \times 10^{-12}$, which we consider 
as a conservative upper bound. This estimate implies that the CP conserving 
part is essentially negligible for $B(K_L\to\pi^0 e^+ e^-)$.

The new measurement of $B(K_S\to\pi^0 e^+ e^-)$ has enabled us to fix
the size of the indirect CP-violating component in $K_L\to\pi^0 e^+ e^-$.
We have also provided arguments that strongly indicate a {\it positive}
interference with the direct CP-violating amplitude. Combining both
contributions we predict 
\begin{equation}
B(K_L\to\pi^0 e^+ e^-)= (3.2^{+1.2}_{-0.8})\times 10^{-11}
\end{equation}
in the Standard Model. The largest contribution is indirect CP-violation,
but direct CP-violation is also substantial and amounts to
$\sim 40\% $ of the rate, mostly from interference of the two components. 
The sizable uncertainty which affects at present
this prediction reflects the poor experimental knowledge 
of the  $K_S \to\pi^0 e^+ e^-$ transition. With more precise 
data on the latter, the theoretical error on $B(K_L\to\pi^0 e^+ e^-)$ 
could in principle be reduced below the $10\%$ level.
In this perspective, we stress that a model-independent confirmation
of the positive interference between direct and indirect CP-violating
amplitudes, which could in principle be obtained by means of lattice QCD,
would also be very useful.

As we have briefly discussed, the total rate is not the only 
interesting observable in this channel: the Dalitz plot 
analysis and the $K_L$--$K_S$ interference, 
in time-dependent distributions, could both be very useful 
in order to extract the direct CP-violating component 
of the $K_L\to\pi^0 e^+ e^-$ amplitude. 

In summary, our analysis demonstrates that the rare decay $K_L\to\pi^0 e^+ e^-$,
despite its complexity, can be well described exploiting additional
experimental input. It implies, in particular, that its branching
fraction is larger than might have been anticipated and has
a substantial sensitivity to direct CP violation and New Physics
in $\Delta S=1$ transitions. If the experimental challenges 
posed by a measurement of $B(K_L\to\pi^0 e^+ e^-)$ can be overcome,
this decay will provide us with most valuable information on
quark flavour physics.

\section*{Acknowledgments}
We thank Augusto Ceccucci for interesting discussions 
and we acknowledge his successful effort in keeping 
secret the NA48 result on $B(K_S \to \pi^0 e^+ e^-)$
till the public announcement. We are grateful
also to Laurie Littenberg, Mara Martini,  
and Ivan Mikulec for useful conversations.
The work of G.D and G.I. is partially supported by IHP-RTN, 
EC contract No.\ HPRN-CT-2002-00311 (EURIDICE).

\section*{Appendix: $\cO(p^6)$ local contributions
to $K^0 \to (\pi^0) \gamma \gamma$}
\addcontentsline{toc}{section}{Appendix: $\cO(p^6)$ 
local contributions to $K^0 \to (\pi^0) \gamma \gamma$}
\label{app:counter}
\newcounter{zahler}
\renewcommand{\theequation}{\Alph{zahler}.\arabic{equation}}
\setcounter{zahler}{1}
\setcounter{equation}{0}

Recent precise data  by NA48 
\cite{NA48_KLpgg} and KTeV \cite{KTeV_KLpgg}, and  some recent theoretical papers 
\cite{GV, Gabbiani:2002bk}, have raised an interesting discussion 
about the determination of the ${\cal O}(p^6)$ local contributions 
to  $K_L \to \pi^0 \gamma \gamma$ and the related role of vector-meson 
contributions~\cite{Littenberg:2002um}. As we shall show 
in the following, a new important piece of information 
in this respect is provided by the very precise measurement 
of the $K_S \to \gamma \gamma$ rate \cite{NA48_KSgg}.
Using chiral symmetry, we can relate counterterm  
contributions to $K_L\to\pi^0 \gamma \gamma$ and $K_S \to \gamma \gamma$
amplitudes. This leads to a substantial reduction of the allowed 
parameter space for the $\cO(p^6)$ couplings and, as a consequence,
to a better understanding of the  vector-meson-dominance (VMD) ansatz.

The $(8_L,1_R)$ weak chiral Lagrangian of $\cO(p^6)$
contains a huge number of operators. However, 
as long as we are interested only in 
$K_L\to\pi^0 \gamma \gamma$ and $K_S \to \gamma \gamma$
decays, we can restrict the attention only to three independent 
combinations \cite{CEP93}.
In particular, we can parameterize all the 
contributions in terms of the following simplified 
effective Lagrangian\footnote{~The strongest simplification 
arises from the fact that we are dealing only with neutral fields, 
which commute with the charge matrix $Q$. For this reason, 
the latter is not explicitly inserted  
in the effective operators in (\ref{eq:p6ctm}).}
\beq
{\cal L}_6 = 
 \frac{G_{8} \alpha }{ 4\pi }\left(  
  a_1  F_{\mu \nu}  F^{\mu \nu} \langle \Delta  \chi_+  \rangle 
+ a_2  F_{\mu \nu}  F^{\mu \nu} \langle \Delta u^\mu u_\mu \rangle 
+ a_3  F_{\mu \lambda}  F^{\mu \sigma} \langle \Delta \{ u^\lambda, u_\sigma \} \rangle 
\right)~,
\label{eq:p6ctm}
\eeq
where, following the notation of Ref.~\cite{Ecker},
we define $\Delta = u \lambda_6 u^\dagger $, 
$u_\mu = i u^\dagger D_\mu U u^\dagger $ and 
$ \chi_+ = u^\dagger {\cal M} u^\dagger + u {\cal M} u$, 
with $U=u^2=e^{i\sqrt{2}\phi/F_\pi}$ 
($\phi$ denotes the pseudoscalar-meson field), $F_\pi \approx$ 93~MeV
and ${\cal M}={\rm diag}(m_\pi^2,m_\pi^2,2m_K^2-m_\pi^2)$. The first 
operator in (\ref{eq:p6ctm}) contributes both to 
$K_L\to\pi^0 \gamma \gamma$ and $K_S \to \gamma \gamma$,
whereas the other two can contribute only 
to $K_L\to\pi^0 \gamma \gamma$.

Historically, the $\cO(p^6)$ local contributions to 
$A$ and $B$ amplitudes of $K_L\to\pi^0 \gamma \gamma$
are parameterized in terms of the three unknown coefficients 
$\alpha_1, \alpha_2$ and $\beta$, defined by \cite{CEP93}
\begin{equation}
A_{\rm CT}=\alpha_1 (z-r_\pi^2)+\alpha_2, \quad B_{\rm CT}=\beta 
\label{eq:ABct}.
\end{equation}
Using the Lagrangian (\ref{eq:p6ctm}) we find 
\beqa
\alpha_1 &=& \frac{m_K^2}{F^2_\pi}( 4 a_2 + 2 a_3 )~,  \no \\
\alpha_2 &=& \frac{m_K^2}{F_\pi^2} \left[ 8 a_1 -4 a_2 + 2 a_3 \right] -0.65~,\no \\
\beta & =& \frac{m_K^2}{F_\pi^2} (-4 a_3) \label{eq:calf} -0.13~, 
\label{eq:KLgg_CT}
\eeqa
where the extra numerical pieces in $\alpha_2$ and $\beta$ are 
residual polynomial parts of the loop amplitudes, renormalized
in the minimal subtraction scheme at the scale $\mu=m_\rho$ \cite{CEP93}. 
It is then easy to check that the VMD ansatz of Ref.~\cite{EPR90}
corresponds to the following choice for the $a_i$:
\begin{equation}
a_1=0 \qquad \frac{m_K^2}{F_\pi^2} a_3 = - \frac{m_K^2}{F_\pi^2} a_2 = 2 a_V 
\label{eq:ABVMD}.
\end{equation}   

The local contributions in (\ref{eq:ABct}) have to be added to the 
unitarized loop amplitudes to obtain the full result. In the case
of the $B$ amplitude this leads to \cite{CEP93}
\beq
B(z)= \beta  +  c\times \left\{  \frac{ 4r_{\pi}^{2}}{z} 
F\left( \frac{z}{r_{\pi }^{2} }\right) 
+ \frac{2}{3} \left( 10 - \frac{z}{r_{\pi }^{2} } \right) 
\left[ \frac{1}{6}+ 
R\left( \frac{z}{r_{\pi }^{2}} \right) \right] +\frac{2}{3} \log\left( \frac{m_\pi^2}{m_\rho^2}
\right) \right\}~,
\label{eq:byz6}
\eeq
where $c$ is the coefficient, determined from 
$K\rightarrow 3\pi$ quadratic slopes, which rules the 
strength of unitarity corrections
(in the numerical analysis we shall employ the value $c=1.1$) and 
the explicit expression of $F(z)$ and $R(z)$ can be found in \cite{CEP93}.
Taking into account that  $F(z)\to - z/12$ and $R(z) \to z/60$ 
for $z\rightarrow 0$, it follows that $B(0)=-1.71+\beta$. 
As shown in Eqs.~(\ref{eq:B110})--(\ref{eq:B111}), the magnitude 
of $B(0)$ is determined in a model-independent way by the 
measurement of $B(K_L \to \pi^0 \gamma\gamma)$ at low diphoton invariant mass. 
Starting from the NA48 result
\beq
 B(K_L \to \pi^0 \gamma\gamma)_{ M_{\gamma \gamma} \in [30-110]~{\rm MeV},\  |y| \in [0-0.2] }
~ <  0.6 \times 10^{-8} \quad{\rm at }\quad  90\, \% {\rm C.L.}, 
\label{eq:NA48_low}
\eeq
which under the assumption of a constant $B$ amplitude becomes
\beq
 B(K_L \to \pi^0 \gamma\gamma)_{  M_{\gamma \gamma}  < 110~{\rm MeV}  }
~ <   0.9 \times 10^{-8} \quad{\rm at }\quad  90\, \% {\rm C.L.},
\eeq
we can derive a first clear bound on the $\cO(p^6)$ counterterms, namely 
\beq
-0.4 < \beta < 3.8~, \qquad   -1.0 < \frac{m_K^2}{F_\pi^2} a_3  < 0.07~.
\label{eq:betaf}
\eeq

A more precise constraint is obtained by means of $B(K_S \to \gamma\gamma)$.
In this case most of the unitarity corrections are implicitly taken into account 
by the overall coupling $G_8$, extracted from the measured 
$K_S \to \pi^+\pi^-$ rate.
Non-trivial $\cO(p^6)$ terms are expected only  
from the (non-VMD) local contributions proportional to $a_1$,  
and from the tiny unitarity corrections associated to the 
process $K_S \to \pi^0 \pi^0 \to \gamma\gamma$ \cite{KH94}.
The recent precise measurement of NA48, 
$B(K_S\to \gamma \gamma ) =(2.78\pm 0.072) \times 10^{-6}$
\cite{NA48_KSgg},
is substantially higher than the $\cO(p^4)$ prediction,
$B(K_S\to \gamma \gamma )^{(4)}=2.1 \times 10^{-6}$ \cite{DEG},
indicating for the first time the need for these $\cO(p^6)$ terms.
Summing the local term generated by  ${\cal L}_6$
to the leading loop amplitude of $\cO(p^4)$ \cite{DEG} we can write
\beq
A(K_S\rightarrow {\gamma}  {\gamma}) =
-i  {2 \alpha G_8  F_\pi \over \pi m^{2}_{K} } ({M}^{2}_{K} - {M}^{2}_{\pi} ) 
\left[ F\left(\frac{1}{r_\pi^2} \right) + \frac{2 m_K^2}{F_\pi^2} a_1 \right]
\epsilon_{1\mu} \epsilon_{2\nu} \left[
  q_1^\nu q_2^\mu - (q_1 q_2) g^{\mu\nu} \right]~.
\label{eq:ksgga} 
\eeq
where the relative sign between local and non-local terms 
has been unambiguously fixed by the sign convention adopted 
in (\ref{eq:KLgg_CT}). Using  $a_1$ as a free parameter to 
fit the experimental branching ratio 
we finally obtain 
\beq
\frac{ 8 m_K^2}{F_\pi^2} a_1 =  (\alpha _1 + \alpha _2 + \beta) + 0.78 =
 1.0 \pm 0.3~.
\label{eq:c1fix}
\eeq
The error in (\ref{eq:c1fix}) is not due to the experimental 
uncertainty, but is a theoretical estimate 
of possible subleading terms: the $\pm 1 \sigma$ 
value in (\ref{eq:c1fix}) corresponds to the following 
two cases: i) Eq.~(\ref{eq:ksgga}) only; ii) Eq.~(\ref{eq:ksgga})
plus the absorptive contribution due to 
$K_S \to \pi^0 \pi^0 \to \gamma\gamma$.

Taking into account the constraints in (\ref{eq:c1fix}) and 
(\ref{eq:betaf}) we are basically left with a single free
parameter to fit both rate and high diphoton invariant mass 
spectrum of $K_L \to \pi^0 \gamma\gamma$. Apparently, the 
constraint on $\beta$ in (\ref{eq:betaf}) is not very 
stringent; however, this does not represent a serious
limitation. The situation becomes much 
more clear if we adopt the $S$--$B$ basis 
for the $K_L \to \pi^0 \gamma\gamma$ amplitude
($S=A+B$):  in this framework the bound in  (\ref{eq:betaf})
tells us that the $B$ amplitude plays a very marginal role
-- but for very low diphoton invariant masses (see Section \ref{sect:CPC}) --
and that both rate and high diphoton invariant 
mass spectrum of $K_L \to \pi^0 \gamma\gamma$ are completely 
dominated by $S$. The latter depend only on two independent 
counterterm combinations, namely $\alpha_1$ and $(\alpha_2+\beta)$,
whose sum is severely constrained by (\ref{eq:c1fix}). 
It is then clear that we are left with a single effective 
coupling, which we chose to be $\alpha_1$. 
Summing the local contributions to the unitarized 
loop amplitude and fitting the NA48 result 
$B(K_L\to \pi^0 \gamma \gamma )=(1.36 \pm 0.05) \times 10^{-6}$~,
we find 
\beq
\alpha_1 = \frac{m_K^2}{F^2_\pi}( 4 a_2 + 2 a_3 ) = 3.4 \pm 0.4~, 
\label{eq:alpha1_fix}
\eeq
where the error reflects mainly the uncertainty  
in (\ref{eq:c1fix}).\footnote{~The central value 
in (\ref{eq:alpha1_fix}) has been obtained with 
the inclusion of the absorptive contribution 
due to $K_L \to 3 \pi^0 \to \pi^0 \gamma\gamma$ 
in the non-local $A$ amplitude \cite{KH94}:
with this choice, we find that the predicted 
$K_L \to \pi^0 \gamma\gamma$ spectrum 
is in excellent agreement with the
observation of NA48 \cite{NA48_KLpgg}. 
If the absorptive contribution 
due to $K_L \to 3 \pi^0 \to \pi^0 \gamma\gamma$ is not included,
the central value of $\alpha_1$ fitted from the rate 
turns out to be about $20\%$ higher, but the resulting 
diphoton spectrum is not in very good agreement 
with NA48 results.  
We have also explicitly checked that the 
central value in (\ref{eq:alpha1_fix})
is completely insensitive to the choice of $\beta$,
varied within the interval (\ref{eq:betaf}).} 

\medskip

A detailed discussion of the results in (\ref{eq:betaf}),
(\ref{eq:c1fix}) and (\ref{eq:alpha1_fix}) goes beyond 
the scope of this work and will be presented elsewhere.  
The quoted figures should also be taken with some care, 
since they are not based on a complete $\chi^2$ analysis. 
On the other hand, we can already draw a few interesting conclusions:
\begin{itemize} 
\item{}
The local structure of $K_S \to \gamma\gamma$ 
and $K_L \to \pi^0 (\gamma\gamma)_{J=0}$ amplitudes
is strongly constrained  by chiral symmetry. Taking 
into account this theoretical constraint and the recent precise 
data by NA48, we are able to fix in a precise way the 
two independent combinations of counterterms, $a_1$ and  $2 a_2 + a_3$,
which control these amplitudes. Interestingly,
these results rule out completely 
or strongly constrain most of the solutions
proposed in \cite{Gabbiani:2002bk}, taking into 
account $K_L \to \pi^0 \gamma\gamma$ data only. 
\item{}
As discussed in Section \ref{sect:CPC}, 
the smallness of the $K_L \to \pi^0 (\gamma\gamma)_{J=2}$ 
amplitude is a direct consequence of 
the experimental result in (\ref{eq:NA48_low}).
We thus agree with the conclusion of 
Ref.~\cite{Gabbiani:2002bk} that the estimate of 
the CPC contribution to $K_L \to \pi^0 e^+ e^-$ 
does not depend on specific assumptions
about $K_L \to \pi^0 \gamma\gamma $ counterterms. 
Since the low diphoton invariant mass analysis of NA48  
contradicts previous findings by KTeV \cite{KTeV_KLpgg},
an independent confirmation of (\ref{eq:NA48_low})
would be very welcome. 
\item{} 
The evidence for a non-vanishing $a_1$ in 
(\ref{eq:c1fix}) shows that the VMD ansatz 
(\ref{eq:ABVMD}) is not exactly fulfilled. 
Nonetheless, VMD contributions are likely 
to provide the dominant $\cO(p^6)$ contribution. 
This statement is not very quantitative 
at the moment due to the uncertainty on 
$a_3$ in (\ref{eq:betaf}). However, we note 
that the values for $a_3$ preferred by the low 
diphoton invariant mass analysis of NA48 are 
those close to the boundaries of (\ref{eq:betaf}):
$(m_K^2/F_\pi^2) a_3 \approx  - 0.9$ or $0.1$.
In other words, the $B$ amplitude is likely to be just 
below the exclusion limit implied by  (\ref{eq:NA48_low}).
If $(m_K^2/F_\pi^2)a_3 \approx  - 0.9$, 
taking into account also the constraint (\ref{eq:alpha1_fix}), 
we would conclude that 
\beqa
\frac{m_K^2}{F_\pi^2} (a_3+a_2) & \approx &  +0.4 \quad \stackrel{ {}_{\rm VMD} }{\longrightarrow} \quad  0~, \\
\frac{m_K^2}{F_\pi^2} (a_3-a_2) & \approx &  -2.2 \quad \stackrel{ {}_{\rm VMD} }{\longrightarrow} \quad  4 a_V~,
\eeqa
which can be interpreted as a clear manifestation of 
vector dominance. 
\end{itemize}

\frenchspacing

\end{document}